\documentclass[aps,pra,twocolumn,showpacs,floatfix,twoside,superscriptaddress]{revtex4-2}

\usepackage{float}
\usepackage{graphicx}
\usepackage{dcolumn}
\usepackage{bm}
\usepackage{physics}
\usepackage{verbatim}
\usepackage{tikz}
\usepackage{wrapfig}
\usepackage{blkarray, bigstrut}
\usepackage{xparse}
\usepackage{xcolor}
\usepackage{amsmath}
\usepackage{amssymb}
\usepackage{bbold}
\usepackage{blkarray}
\usepackage{mathdots}
\usepackage{bm}
\usepackage{hyperref}
\usepackage{caption}
\usepackage{mathrsfs}
\hypersetup{
    colorlinks=true,
    linkcolor=blue,
    filecolor=magenta,      
    urlcolor=cyan,
    pdftitle={Overleaf Example},
    pdfpagemode=FullScreen,
    }

\newcommand{\BL}{\mathcal{L}}
\newcommand{\BP}{\mathcal{P}}
\newcommand{\BT}{\mathcal{T}}
\DeclareMathOperator{\im}{Im}
\DeclareMathOperator{\re}{Re}

\begin{document}

\preprint{APS/123-QED}

\title{Multi-block exceptional points in open quantum systems}

\author{Aysel Shiralieva}
\author{Grigory A. Starkov}
\author{Björn Trauzettel}

\affiliation{Institute for Theoretical Physics and Astrophysics,
University of Würzburg, D-97074 Würzburg, Germany}
\affiliation{Würzburg-Dresden Cluster of Excellence ctd.qmat, Germany}

\date{\today}

\begin{abstract}
Open quantum systems can be approximately described by non-Hermitian Hamiltonians (NHHs) and Liouvillian superoperators. The two approaches differ by quantum jump terms corresponding to a measurement of the system by its environment. We analyze the emergence of exceptional points (EPs) in NHHs and Liouvillian superoperators. In particular, we show how EPs in NHHs relate to a novel type of EPs --- multi-block EPs --- in the no-jump Liouvillian, i.e. the Liouvillian superoperator in absence of quantum jump terms. We further analyze how quantum jump terms modify the multi-block structure. To illustrate our general findings, we present two prime examples: qubits and qutrits coupled to additional ground state levels that serve as sinks of the population. In those examples, we can navigate through the EP block structure by a variation of physical parameters. We analyze how the dynamics of the population of the states is affected by the order of the EPs. Additionally, we demonstrate that the quantum geometric tensor serves as a sensitive indicator of EPs of different kinds. 
\end{abstract}

\maketitle

\section{\label{sec:introduction}Introduction}

Non-Hermitian systems are characterized by a spectrum with complex eigenvalues. This complexity enables these systems to exhibit spectral singularities known as exceptional points (EPs), where both eigenvalues and eigenvectors of the corresponding evolution operator coalesce. The presence of EPs gives rise to a range of intriguing phenomena that do not occur in Hermitian systems and that can have promising practical applications~\cite{Ashida02072020}. For example, they can be used to enhance sensor sensitivity~\cite{Hodaei2017,Wiersig:20sensors} or for adiabatic state preparation~\cite{PhysRevA.102.040201}. They are also particularly relevant for certain types of topological phase transitions~\cite{gong_topological_2018,RevModPhys.93.015005,Okuma2023}. EPs have been experimentally realized in a number of physical platforms, including dielectric resonator chains \cite{Poli2015}, optical waveguides \cite{PhysRevLett.115.040402,Weimann2017,Cerjan2019}, optical microcavities \cite{chen_exceptional_2017}, topological insulator lasers \cite{Bandres2018,Klembt2021}, photonic crystals \cite{Zhou2018}, topolectric circuits \cite{Helbig2020}, and plasmonics \cite{Plasmonics_EP_PhysRevB.94.201103} (see Ref.~\cite{Wang:23} for a recent review on photonic platforms).

Non-Hermiticity naturally arises in the context of open quantum systems. Under common assumptions, like Born and Markov approximations, the dynamics of the reduced density matrix of the system can be described by the Lindblad master equation. 
Within this framework, the Liouvillian superoperator, capturing the dynamics of the system, can be viewed as a composition of two terms: one describing coherent non-unitary dynamics associated with an effective non-Hermitian Hamiltonian (NHH), and another one describing the non-coherent influence of quantum jumps attributed to the continuous measurement of the system by its environment. The former term, the part of the Liouvillian superoperator excluding the quantum jump term, we call no-jump Liouvillian.

\begin{table}[t]
\caption{List of abbreviations used in the article and corresponding full names. Notation follows Ref.~\cite{minganti_quantum_2019}.}
\label{tab:abbreviations}
\centering
\begin{tabular}{l l} 
\hline\hline
Full name & Abbreviation \\
\hline
Non-Hermitian Hamiltonian                    & NHH \\
Liouvillian                                  & $\mathcal{L}$ \\
No-jump Liouvillian                          & $\mathcal{L}^\prime$ \\
Liouvillian of the effective system          & $\mathcal{L}_\mathrm{eff}$ \\
No-jump Liouvillian of the effective system  & $\mathcal{L}^\prime_\mathrm{eff}$ \\
Exceptional point                            & EP \\
Diabolical point                             & DP \\
Hamiltonian exceptional point                & HEP \\
Liouvillian exceptional point                & LEP \\
LEP of no-jump Liouvillian                  & njLEP \\
Quantum geometric tensor                     & QGT \\
\hline\hline
\end{tabular}
\end{table}

We are particularly interested in the structure of EPs in NHHs, no-jump Liouvillians, and Liouvillian superoperators. In their seminal work, Minganti et al.~\cite{minganti_quantum_2019} compared EPs of NHHs (HEPs), no-jump Liouvillians (njLEPs), and Liouvillian superoperators (LEPs), concluding that they can substantially differ from each other. What is missing, however, is a proper characterization of this difference, in particular, in the Jordan block basis of no-jump Liouvillians and Liouvillian superoperators, where the difference becomes most pronounced. 

The study of LEPs is of significant interest, and their properties in dissipative superconducting qubits have been studied experimentally~\cite{chen_decoherence-induced_2022}.

In our work, we prove a general relation between HEPs, njLEPs and LEPs. We show in particular that HEPs become multi-block EPs (also called derogatory in the mathematical literature) in the corresponding no-jump Liouvillian eigenspectrum. They are characterized by multiple Jordan blocks corresponding to the same eigenvalue. To be exact, an $n$-th order HEP results in a multi-block EP with block sizes $(2n-1),\ (2n-3),\dotsc,1$ in the Liouvillian superoperator. Note that our result refines the previous conjecture that an $n$-th order HEP corresponds to a $(2n-1)$-th order LEP ~\cite{Jan_Wiersig_EP_PhysRevA.101.053846}. In fact, this conjecture misses the multi-block structure of the LEPs. Notably, multi-block EPs have recently been discussed in the context of NHHs (not Liouvillian superoperators like in our case) under the name of Fragmented EPs~\cite{bid_PhysRevResearch, bid_2025}. 

If we add dissipation terms to the problem, that can be (but not have to be) of quantum jump type, the multi-block EPs of the no-jump Liouvillian or Liouvillian superoperator split up completely or partially into subblocks. Interestingly, the character of the splitting is directly influenced by the type of EP and its particular block structure. Typically, the order of the EPs is reduced by the dissipation terms, but some EPs are stable against certain perturbations. We illustrate these general findings with concrete examples, in particular effective non-Hermitian qubit (qutrit) systems that arise from projecting out the ground state of a larger dissipative three- (four-)level system (see Sec.~\ref{sec:the model} for a detailed derivation). We show how the population dynamics of these levels is affected by the presence of EPs. Furthermore, we demonstrate that EPs (of different kinds) can be identified as singularities in the quantum geometric tensor (QGT), which quantifies the geometric distance of quantum states in parameter space.

The article is organized as follows. In Sec.~\ref{sec:the n-level model}, we introduce the effective \((N-1)\)-level model and establish a general relationship between the order of EPs in NHHs and those in the corresponding no-jump Liouvillians. In Sec.~\ref{sec: examples}, we exemplify this relationship by effective non-Hermitian qubit and qutrit models. In Sec.~\ref{sec:numerical results}, we illustrate the dynamical evolution of qubit and qutrit level populations close to EPs. In Sec.~\ref{sec:QGT}, we demonstrate how the QGT can be employed to detect EPs in parameter space. We conclude in Sec.~\ref{sec:con} and describe details of the calculations in the Appendix.

\section{\label{sec:the n-level model}General model}
\subsection{Effective $(N-1)$-level subsystem in the Hybrid-Liouvillian formalism \label{sec:the model}}

\begin{figure}[t]
\centering
\includegraphics[width=.8\linewidth]{ 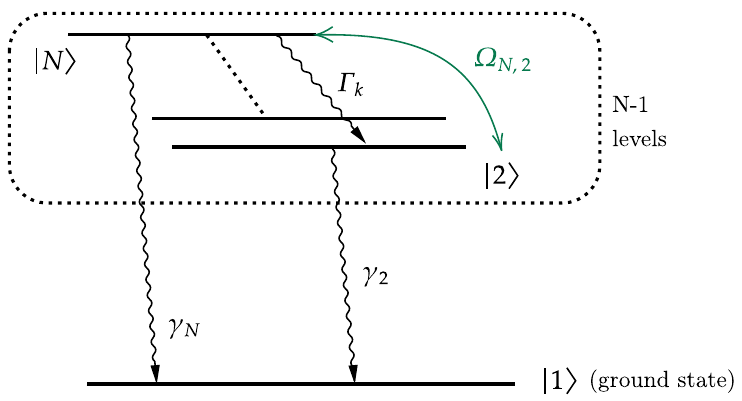}
    \caption{Illustration of the \(N - 1\) excited levels undergoing non-Hermitian evolution. The $\Omega_{ij}$ values represent coherent drives between pairs of excited states and the $\gamma_{j}$ values denote decay rates from the excited states to the ground state. $\Gamma_k$ parametrize the dissipation terms affecting exclusively the $(N-1)$ excited levels.}
\label{fig:N levels model}
\end{figure}

Let us consider an $N$-level system coupled to the environment, resulting in the dynamics governed by the Lindblad master equation.
\begin{align}
    \dot{\hat\rho} & =-i[\hat{H}, \hat\rho]+\sum_{k} \Gamma_{k} \left(\hat{L}_k \hat\rho \hat{L}_k^{\dagger}-\frac{1}{2}\left\{\hat{L}_{k}^{\dagger} \hat{L}_{k}, \hat\rho\right\}\right) \nonumber \\
    &=-i\left(\hat{H}_{\text {nh}} \rho-\rho \hat{H}_{\text {nh}}^{\dagger}\right)+\sum_{k} \Gamma_{k} \hat{L}_{k} \hat\rho \hat{L}_{k}^{\dagger} \label{eq:Lindblad}\\
      \hat{H}_{\mathrm{nh}}&=\hat{H}-\frac{i}{2} \sum_{i\neq j} \Gamma_{k} \hat{L}_{k}^{\dagger} \hat{L}_{k}.\label{eq:Heff and Lindblad equation}
\end{align}
Here, $\hat H$ is the Hamiltonian and $\hat\rho$ is the reduced density matrix of the system.
The jump operators $\hat L_k$ describe transitions between the levels of the system induced by the coupling to the environment, that occur with rates $\Gamma_k$. For simplicity, all parameters are taken to be real. We schematically illustrate the system in Fig.~\ref{fig:N levels model}.

The Lindblad master equation~\eqref{eq:Lindblad} can be rewritten as
\begin{equation}
    \dot{\hat\rho} = \BL \hat\rho,
\end{equation}
where $\BL$ is the linear Liouvillian superoperator acting on the reduced density matrix. To reveal the matrix form of $\BL$, we vectorize the density operator. By vectorization we mean the representation, in which a generic operator $\hat{\chi} = \sum_{m, n} c_{m, n} |m\rangle \langle n|$
is identified with a vector $\vec{\chi} = \sum_{m, n} c_{m, n} |m\rangle \otimes \left| n^* \right\rangle$, given an orthonormal basis of the Hilbert space $\{|n\rangle\}$ (see Appendix of Ref.~\cite{minganti_hybrid-liouvillian_2020} for more details). In this vectorized representation, the matrix of the Liouvillian superoperator can be written explicitly as
\begin{multline}
    \BL  = -i \qty ( \hat{H} \otimes \mathbb{1} - \mathbb{1} \otimes \hat{H}^{\text{T}} ) +\\
     \sum_{k} \Gamma_k \qty[\hat{L}_{k} \otimes \hat{L}^*_{k}- \frac{\hat{L}_{k}^{\dagger} \hat{L}_{k}  \otimes \mathbb{1}  + \mathbb{1} \otimes \hat{L}^{T}_{k} \hat{L}^*_{k}}{2}] = \\
    \qty[\qty ( -i \hat{H}_{\text {nh }} )  \otimes \mathbb{1}  + \mathbb{1} \otimes \qty ( i \hat{H}_{\text {nh }}^* )] + \sum_{k} \Gamma_{k} \hat{L}_{k} \otimes \hat{L}^*_{k} \label{eq:superoperator Lindblad}.
\end{multline}
By means of Eq.~\eqref{eq:superoperator Lindblad}, we can check that $\BL$ is indeed non-Hermitian. It also satisfies a generalized $\BP\BT$-symmetry, where $\BT$ is understood as the complex conjugation operator, while $\BP$ swaps two terms in the Kronecker products~\cite{Sa_2023}.

To analyze the structure of LEPs, we adopt the \emph{hybrid-Liouvillian approach}~\cite{minganti_hybrid-liouvillian_2020}. This framework allows us to separate the evolution governed by the NHH (captured by the terms in the square brackets in Eq.~\eqref{eq:superoperator Lindblad}) from the full Lindbladian dynamics, enabling us to compare the HEPs with LEPs. 
The core of this method involves interpolating between the no-jump Liouvillian $\mathcal{L}'$ and the full Liouvillian $\mathcal{L}$ via a tuning parameter. This is structured such that one limit of this parameter recovers the no-jump dynamics, while the other corresponds to the full Liouvillian evolution. Physically, this method represents the (possibly non-ideal) post-selection of quantum trajectories in which no jumps occur~\cite{naghiloo_quantum_2019, chen_quantum_2021}, thereby realizing NHH evolution within the system.
Within this framework, we treat the ground state as the sink for the population of the states of the system and focus on the dynamics of the $\left(N-1\right)$ excited states.

In our choice of system-bath coupling, 
the first $l$ quantum jump terms connect only the excited states, see Fig.~\ref{fig:N levels model}, while the rest describe the dissipation from the excited states into the ground state denoted by
\begin{equation}
    \hat L_{l+k-1} = |1\rangle\langle k|,\quad \mathrm{for}\quad 2\leqslant  k\leqslant N.
\end{equation}
To distinguish the corresponding terms, let us denote $\gamma_k = \Gamma_{l+k-1}$ for $2\leqslant k\leqslant N$. For simplicity, we assume that there is no coherent drive between the excited states and the ground state. This way, we can write the system's Hamiltonian explicitly as
\begin{equation}
    \hat H = \sum_{i>1} \Delta_i|i\rangle\langle i| +\sum_{j>i>1} \Omega_{ij} \qty [|i\rangle\langle j| + |j\rangle\langle i|].
\end{equation}
Here, $\Omega_{i,j}$ are the coherent drives between the excited levels, which we assume for simplicity to be real.

If we arrange the components of the \emph{vectorized} density matrix such that the elements corresponding to the evolution within the non-Hermitian \(\qty(N - 1)\)-level system are the first entries, then the Liouvillian superoperator of the full system has block-diagonal form
\begin{equation}\label{eq:Lindbladian-block-diag-form}
\begin{aligned}
     {\mathcal{L}} &= \begin{pmatrix}
        {\mathcal{L}}_{\text{eff}} & \mathbb{0}\\
        {Q} & {G} 
    \end{pmatrix},
\end{aligned}
\end{equation}
Here, the upper-left block,  ${\mathcal{L}}_{\text{eff}}$, represents the dynamics of the $N - 1$ excited levels, which we call from now on \emph{the effective system}. The quantum jumps that occur pairwise between the excited states also enter $\mathcal{L}_{\text{eff}}$.
The lower-right block, ${G}$, represents how the levels of the effective system affect the coherences and the population of the ground state. The block ${Q}$ contains quantum jumps that happen pairwise between the \( \qty ( N - 1 ) \) excited levels and the ground state. The upper-right block is strictly zero because of our assumption that the quantum jumps only connect the excited states to the ground state and not vice versa.

Mathematically, we can trace out the components of the density operator corresponding to the ground state using the projection operator $\check P = \hat P\otimes\hat P$, where $\hat P = \sum_{i>1} |i\rangle\langle i|$. Physically, it can be achieved by post-selection techniques. If we use the same ordering for the basis of vectorized density operator as in Eq.~\eqref{eq:Lindbladian-block-diag-form}, $\check P$ acquires the simple form
\begin{equation}
    \check P = \begin{pmatrix}
                \mathbb{1}_{(N-1)^2\times( N-1)^2} & \mathbb{0}_{(N-1)^2\times(2N-1)}\\
                \mathbb{0}_{(2N-1)\times(N-1)^2} & \mathbb{0}_{(2N-1)\times(2N-1)}
               \end{pmatrix},
\end{equation}
where the sizes of the blocks are the same as in Eq.~\eqref{eq:Lindbladian-block-diag-form}.
Due to the special structure of the Liouvillian superoperator matrix, $\check P$ and $ \BL$ commute and
\begin{equation}
    \check P \hat\rho(t) = \check P e^{ \BL t}\rho_0 = \qty(\check P e^{ \BL t} \check P) \qty(\check P\rho_0),\label{eq:rho-proj-dynamics}
\end{equation}
where
\begin{equation}
    \check P e^{ \BL t} \check P = \begin{pmatrix}
        e^{\BL_\mathrm{eff}t} & \mathbb{0}\\
        \mathbb{0} & \mathbb{0}
    \end{pmatrix}\label{eq:proj-exponential}
\end{equation}
Hence, under the assumption that the quantum jumps only connect the excited states to the ground state and not vice versa, $\BL_\mathrm{eff}$ indeed describes the dynamics of the subsystem composed of the excited states. The explicit form of $\BL_\mathrm{eff}$ is given by
\begin{align}
\BL_\mathrm{eff} &=
\left[ \left( -i \hat{H}_{\text{eff}} \right) \otimes \hat{\mathbb{1}} + \hat{\mathbb{1}} \otimes \left( i \hat{H}_{\text{eff}}^* \right) \right] + \sum_{k=1}^{l} \Gamma_k \, \hat{L}_k \otimes \hat{L}_k^*,\label{eq:eff_Lindbladian}\\
\hat H_\mathrm{eff} &= \sum_{j>1} (\Delta_j - i\gamma_j/2)|j\rangle\langle j| + \sum_{j>i>1} \Omega_{ij} \qty(|i\rangle\langle j| + |j\rangle\langle i|)\nonumber \\
& - i\sum_{k=1}^{l} \Gamma_k \hat L_k^\dagger \hat L_k.
\end{align}
Note that the rates $\gamma_j$ enter only into $\hat H_\mathrm{eff}$. In this sense, $\hat H_\mathrm{eff}$ can be tuned to some extent independently of the quantum jump terms, which depend exclusively on $\Gamma_k$.


\subsection{Relation between HEPs and the corresponding njLEPs}\label{sec: NHH EPs and LEPs }

The square bracket term in Eq.~\eqref{eq:eff_Lindbladian} describes the coherent evolution with the effective NHH $\hat H_\mathrm{eff}$. We denote it as
\begin{equation}
\mathcal{L}^\prime_{\mathrm{eff}} =
\left( -i \hat{H}_{\text{eff}} \right) \otimes \hat{\mathbb{1}} + \hat{\mathbb{1}} \otimes \left( i \hat{H}_{\text{eff}}^* \right) \label{eq:eff_Liouvillian}
\end{equation}
Its physical meaning can be understood, if we write an effective Schr\"odinger equation for a pure state $|\psi(t)\rangle = \sum_{i>1} c_i(t)|i\rangle$ of the effective subsystem:
\begin{equation}
    i \frac{d}{dt} |\psi(t)\rangle = \hat H_\mathrm{eff}|\psi(t)\rangle.\label{eq:eff_schroedinger}
\end{equation}
Differentiating the corresponding density operator $\hat\rho_\psi(t)=|\psi(t)\rangle\langle\psi(t)|$ with the help of Eq.~\eqref{eq:eff_schroedinger}, we obtain 
\begin{equation}
    \dot{\hat\rho}_\psi = \BL^\prime_\mathrm{eff}\hat \rho_\psi
\end{equation}
Therefore, $\BL^\prime_\mathrm{eff}$ has the same relation to the effective NHH $\hat H_\mathrm{eff}$ as the quantum Liouville equation to the Hamiltonian $\hat H$ in the Hermitian case. Following this reasoning, we refer to $\BL^\prime_\mathrm{eff}$ as effective no-jump Liouvillian.

The effective Liouvillian superoperator differs from the effective no-jump Liouvillian by the presence of the quantum jump terms $\Gamma_k \hat L_k\otimes \hat L_k^*$. Therefore, the dynamics due to an effective NHH description substantially differs from the Liouvillian superoperator dynamics. However, it is possible to establish a general correspondence between the spectra of the effective no-jump Liouvillian and the corresponding NHH, which we discuss in the rest of the subsection. The quantum jump terms can then be analyzed as perturbations of the effective no-jump Liouvillian.

Firstly, let us assume that $\hat H_\mathrm{eff}$ is diagonalizable. If $|u\rangle$ and $|v\rangle$ are two right eigenvectors of $\hat H_\mathrm{eff}$ corresponding to the eigenvalues $\varepsilon_u$ and $\varepsilon_v$, then in the vectorized notation $|u\rangle\otimes|v^*\rangle$ is the eigenvector of $\BL^\prime_\mathrm{eff}$ corresponding to the eigenvalue $\lambda_{u,v} = i(\varepsilon_v^* - \varepsilon_u)$.

Secondly, let us consider the case, where $\hat H_\mathrm{eff}$ of dimension $n$ has an $n$-th order EP corresponding to the eigenvalue $\varepsilon$. This means that $\hat H_\mathrm{eff}$ is similar to the $n\times n$ Jordan matrix $J_n(\varepsilon)$:
\begin{align} 
    J_n(\varepsilon) &= \begin{pmatrix}
        \varepsilon & 1 & 0 & 0 & 0 & \dotsb & 0\\
        0 & \varepsilon & 1 & 0 & 0 &  \dotsb & 0\\
        0 & 0 & \varepsilon & 1 & 0 & \dotsb & 0 \\
        \vdots & \ddots & \ddots & \ddots & \ddots & \ddots & \vdots\\
        0&\dotsb&0& 0& \varepsilon& 1&0\\
        0 & \dotsb & 0 & 0 & 0 & \varepsilon & 1 \\
        0 & \dotsb & 0 & 0 & 0 & 0 & \varepsilon \\
    \end{pmatrix},\\
    \hat H_\mathrm{eff} &= S J_n(\varepsilon)S^{-1},\label{eq:similarity}
\end{align}
where $S$ is some invertible matrix describing a generalized basis transformation. In the following, we denote the similarity between the matrices $\sim$. As such, $\hat H_\mathrm{eff}\sim J_n(\varepsilon)$ is equivalent to Eq.~\eqref{eq:similarity}. If the Jordan vectors of $\hat H_\mathrm{eff}$ are known (columns of basis matrix $S$), the Jordan basis of $\BL^\prime_\mathrm{eff}$ can be constructed explicitly using the general procedure which we explain in Appendix~\ref{sec: app A EPs Lindblad and NHH general relation}.
As a byproduct, we also determine the values and order of EPs in the effective no-jump Liouvillian. The result is
\begin{multline} \label{eq: general relation}
    \BL^\prime_\mathrm{eff} = \\\qty(-i\hat H_\mathrm{eff})\otimes\hat{\mathbb 1} + \hat{\mathbb 1}\otimes\qty(i\hat H_\mathrm{eff}^*) \sim
    J_n(-i\varepsilon)\otimes\hat{\mathbb 1} + \hat{\mathbb 1}\otimes J_n(i\varepsilon^*) \\
    \sim J_{2n-1}(\lambda_{\BL^\prime_\mathrm{eff}}) \oplus J_{2n-3}(\lambda_{\BL^\prime_\mathrm{eff}})\oplus \ldots\oplus J_{3}(\lambda_{\BL^\prime_\mathrm{eff}})\oplus J_{1}(\lambda_{\BL^\prime_\mathrm{eff}}),
\end{multline}
where $\lambda_{\BL^\prime_\mathrm{eff}} = i(\varepsilon^*-\varepsilon)=2\im{\varepsilon}$. Note that $\BL^\prime_\mathrm{eff}$ is an $n^2 \cross n^2$ matrix, which is consistent with the total size of all the blocks in the decomposition added together: $ \sum_{l = 1}^{n} (2 l - 1) = n^2$.

This result is quite remarkable. If the effective NHH is tuned to an EP, the corresponding effective no-jump Liouvillian naturally exhibits a novel type of EP characterized by multiple Jordan blocks at the same eigenvalue. Moreover, the size of the largest block exceeds the order of the EP exhibited by $\hat H_\mathrm{eff}$. To some extent, this result has been previously discussed, for instance, in Ref.~\cite{Jan_Wiersig_EP_PhysRevA.101.053846}, where a physical argument was provided that the order of the EP on the level of the effective no-jump Liouvillian should be $(2n-1)$, which coincides with the size of the largest block. However, the multi-block character of the njLEPs has been missed. We also provide the general procedure to construct the Jordan basis of $\BL^\prime_\mathrm{eff}$. This basis can then be used to study the effect of the quantum jump terms.

Let us also consider the case, where the effective Hamiltonian has the general Jordan form
\begin{equation}
    \hat H_\mathrm{eff} \sim \bigoplus_i J_{n_i}(\varepsilon_i)
\end{equation}
and
\begin{multline}
    \BL^\prime_\mathrm{eff} = \qty(-i\hat H_\mathrm{eff})\otimes\hat{\mathbb 1} + \hat{\mathbb 1}\otimes\qty(i\hat H_\mathrm{eff}^*)\\
    \sim \qty( \bigoplus_i J_{n_i}(-i\varepsilon_i))\otimes \hat{\mathbb 1} + \hat {\mathbb 1}\otimes \qty(\bigoplus_j J_{n_j}(i\varepsilon_j^*)).
\end{multline}
Note that the Jordan block sizes are $n_i\geqslant 1$. Analogously to the case of the single EP, the Jordan basis of the effective no-jump Liouvillian can be constructed out of the Jordan basis of $\hat H_\mathrm{eff}$. As a result,
\begin{equation}
    \BL^\prime_\mathrm{eff} \sim \bigoplus_{i,j}\qty [\bigoplus_{k=1}^{\min{(n_i,n_j)}} J_{n_i+n_j-(2k-1)}(\lambda_{i,j})]\label{eq:no-jump-jordan-form}
\end{equation}
with $\lambda_{i,j} = i(\varepsilon_j^*-\varepsilon_i)$. Hence, EPs of the effective Hamiltonian result in a multi-block EP of the effective no-jump Liouvillian.

When the quantum jump terms are added, the multi-block character of the EPs needs to be taken into account to accurately describe the eigenspectrum. For a general perturbation, every block of size $n$ splits like $n$ branches of the complex $n$-th order root, which follows from the Lidskii-Vishik-Lyusternik theorem~\cite{lidskii_perturbation_1966, moro_lidskii--vishik--lyusternik_1997}.

In the next section, we illustrate these findings by two simple examples: non-Hermitian qubits and qutrits.

\section{LEPs for the non-Hermitian qubits and qutrits} \label{sec: examples}

Building on the hybrid-Liouvillian framework introduced in Sec.~\ref{sec:the n-level model}, we now illustrate the general relation derived in Eq.~(\ref{eq: general relation}) by analyzing a non-Hermitian qubit (qutrit) embedded within a three(four)-level system. Furthermore, we use these examples to discuss how the multi-block EPs realized in the spectrum of the effective no-jump Liouvillian get split by the inclusion of the quantum jump terms.

\subsection{Non-Hermitian qubit}

\subsubsection{njLEPs in non-Hermitian qubits}

\begin{figure}[h!]
    \centering
    \includegraphics[width=.6\linewidth]{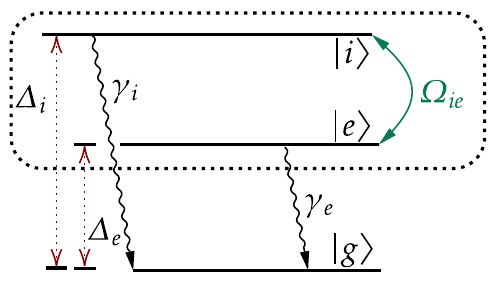}
    \caption{Realization of a non-Hermitian qubit formed by the two upper excited states in a dissipative three-level system.}
    \label{fig:NH qubit}
\end{figure}
We first consider the case of the effective non-Hermitian qubit realized by the two excited states of the three-level system. We denote the three levels as $|g\rangle$, $|e\rangle$ and $|i\rangle$, where $|g\rangle$ stands for the ground state.
The Hermitian part of the Hamiltonian takes form
\begin{equation}
    \hat{H}^{(0)} = \Delta_e \ket{e} \bra{e} + \Delta_i \ket{i} \bra{i} +  \Omega_{ie} \left 
            [  \ket{e} \bra{i}  + \ket{i} \bra{e} \right], \label{eq: H_RWA qubit}
\end{equation}
Here, $\Omega_{ie}$ represents the coupling rate between the excited states from an applied drive with frequency detunings $\Delta_e$ and $\Delta_i$. 
In the absence of the intra-qubit dissipative terms, the effective Hamiltonian of the $|i\rangle-|e\rangle$ qubit is
\begin{equation}
    \hat H_\mathrm{eff} = \begin{pmatrix}
           - \frac{i\gamma_i}{2}  + \Delta_i & \Omega_{ie} \\
            \Omega_{ie} &  - i \frac{\gamma_{e}}{2} + \Delta_e 
            \end{pmatrix},\label{eq: qubit-heff-basic}
\end{equation}
where $\gamma_{i,e}$ are the dissipation rates from the levels $|i\rangle$ and $|e\rangle$ respectively to the ground state (see Section~\ref{sec:the model}).
We assume $\Omega_{ie}$ to be real.

The effective Hamiltonian~\eqref{eq: qubit-heff-basic} exhibits a second-order EP provided
\begin{equation}
    \Delta_i=\Delta_e=\Delta,\qquad (\gamma_i-\gamma_e)^2 = 16\Omega_{ie}^2\neq0.
\end{equation}
The corresponding eigenenergy is $\varepsilon_\mathrm{EP}=\Delta - i(\gamma_i+\gamma_e)/4$.
If $\hat H_\mathrm{eff}$ is tuned to this second-order EP, the effective no-jump Liouvillian acquires the form
\begin{multline}
    \BL_\mathrm{eff}^\prime = \left( -i \hat{H}_{\text{eff}} \right) \otimes \hat{\mathbb{1}} + \hat{\mathbb{1}} \otimes \left( i \hat{H}_{\text{eff}}^* \right)=
    \\
    -\frac{\gamma_i+\gamma_e}{2}\mathbb{1}_{4\times4} + \begin{pmatrix}
        \mp2\Omega_{ie} & i\Omega_{ie} & -i\Omega_{ie} & 0\\
        i\Omega_{ie} & 0 & 0 & -i\Omega_{ie}\\
        -i\Omega_{ie} & 0 & 0 & i\Omega_{ie}\\
        0 & -i\Omega_{ie} & i\Omega_{ie} & \pm2\Omega_{ie}
    \end{pmatrix}-
\end{multline}
It is straightforward to work out the Jordan normal form of $\BL_\mathrm{eff}^\prime$. As a result, one can confirm that it has a four-fold degenerate eigenvalue $\lambda_{\BL^\prime_\mathrm{eff}} = 2\im{\varepsilon_\mathrm{EP}} = -(\gamma_i+\gamma_e)/2$ and
\begin{equation}
    \BL_\mathrm{eff}^\prime\sim J_3(\lambda_{\BL^\prime_\mathrm{eff}})\oplus J_1(\lambda_{\BL^\prime_\mathrm{eff}}).\label{eq: qubit-ep}
\end{equation}

\subsubsection{\label{sec: quantum jumps}Quantum jumps as perturbations of the system}

To study the effect of the quantum jump terms on the multi-block EPs described in Eq.~\eqref{eq: qubit-ep} of the effective no-jump Liouvillian, we add a dissipative term connecting the levels $|i\rangle$ and $|e\rangle$:
\begin{equation}
    \BL_\mathrm{eff} = \BL_\mathrm{eff}^\prime + \Gamma \qty[\hat L\otimes\hat{L}^* - \frac{\hat L ^\dagger\hat L\otimes\mathbb{1} + \mathbb{1}\otimes\hat L^T \hat L^*}{2}]
\end{equation}
We consider two variants of the intra-qubit dissipation:
\begin{enumerate}
\item[(i)] $\hat L = |e\rangle\langle i|$, describing the decay from the second excited state into the first one; and
\item[(ii)] $\hat L = \hat \sigma_y$ --- combined bit- and phase-flip error.
\end{enumerate}

In both cases (i) and (ii), $\Gamma$ is a non-negative coefficient that determines the rates of the non-unitary terms. Specifically, in case (i), it describes the decay rate from the level $\ket{i}$ to $\ket{e}$, whereas in case (ii), it determines the rate of the bit--phase-flip channel. The quantum channels in cases (i) and (ii) are treated as distinct models and are analyzed independently.

\textbullet\ In case (i), the effective Liouvillian superoperator takes the form
\begin{multline}
    \BL_\mathrm{eff}= -\frac{\gamma_i+\gamma_e+\Gamma}{2} \mathbb{1}_{4\times4} +
    \\
    \begin{pmatrix}
        -\frac{(\gamma_i-\gamma_e)+\Gamma}{2} & i\Omega_{ie} & -i\Omega_{ie} & 0\\
        i\Omega_{ie} & 0 & 0 & -i\Omega_{ie} \\
        -i\Omega_{ie} & 0 & 0 & i\Omega_{ie} \\
        \Gamma & -i\Omega_{ie} & i\Omega_{ie} & \frac{(\gamma_i-\gamma_e)+\Gamma}{2}
    \end{pmatrix}.
\end{multline}
The corresponding characteristic equation is of the fourth order. Hence, all the eigenvalues can be found analytically using the Cardano formula:
\begin{equation}
\label{eq:lambda quantum jumps}
\begin{aligned}
    \lambda_k &= -\frac{\gamma_i+\gamma_e+\Gamma}{2}
    + \xi^{k-1}\qty( \nu + \sqrt{ \nu^2 - (\kappa/12)^3 })^{1/3} \\
   & +\frac{\kappa/12}{\xi^{k-1}\qty( \nu + \sqrt{ \nu^2 - (\kappa/12)^3 })^{1/3}},\quad k=1,2,3\\
    \lambda_4 &= -\tfrac{1}{2} (\Gamma + \gamma_{e} + \gamma_{i}),
\end{aligned}
\end{equation}  
where $\nu = \Gamma \Omega_{ie}^2$, $\kappa = (\Gamma+\gamma_i-\gamma_e)^2-16\Omega_{ie}^2$. The parameter $\xi$ is the third-order root of unity $\xi = e^{i\frac{2\pi}{3}} = (-1+i\sqrt{3})/2$.
The effect of the quantum jumps in this case is twofold: It shifts all the eigenvalues by $\Gamma/2$ and splits the Jordan block of size 3.
If the effective no-jump Liouvillian is initially tuned to an EP, $(\gamma_i-\gamma_e)=\pm4\Omega_{ie}$, then $\kappa\approx \pm8\Gamma\Omega_{ie}$ and to leading order in $\Gamma$
\begin{equation}\label{eq: cube_root_splitting_NH_qubit}
    \lambda_{k=1,2,3} \approx -\frac{\gamma_i+\gamma_e}{2} + (2\Omega_{ie}^2)^{1/3} \Gamma^{1/3}\xi^{k-1},
\end{equation}
i.e. they split exactly like three branches of the complex $z^{1/3}$ function, in agreement with the Lidskii-Vyshitski-Lyusternnik theorem ~\cite{lidskii_perturbation_1966,moro_lidskii--vishik--lyusternik_1997}.

\textbullet\ In case (ii), the effective Liouvillian superoperator is
\begin{multline}
    \BL_\mathrm{eff}= -\frac{\gamma_i+\gamma_e+2\Gamma}{2} \mathbb{1}_{4\times4} +
    \\
    \begin{pmatrix}
        -\frac{\gamma_i-\gamma_e}{2} & i\Omega_{ie} & -i\Omega_{ie} & {\Gamma}\\
        i\Omega_{ie} & 0 & -{\Gamma} & -i\Omega_{ie} \\
        -i\Omega_{ie} & -{\Gamma} & 0 & i\Omega_{ie} \\
        {\Gamma} & -i\Omega_{ie} & i\Omega_{ie} & \frac{\gamma_i-\gamma_e}{2}
    \end{pmatrix}
\end{multline}
and its eigenvalues are
\begin{equation}\label{eq:eigenvalues qubit}
    \begin{aligned}
        \lambda_1 &= \frac{- \qty ( \gamma_{i} + \gamma_{e} ) }{2}, \quad \lambda_4 = \frac{- \qty ( 4 \Gamma + \gamma_{i} + \gamma_{e} ) }{2} \\   \lambda_{2,3} &= \frac{- \qty ( 2 \Gamma + \gamma_{i} + \gamma_{e}) \pm \sqrt{ 4 \Gamma^2 + \qty ( \gamma_{i} - \gamma_{e})^2 - 16 \Omega_{ie}^2 }  }{2}
    \end{aligned}
\end{equation}
By tuning the parameters to the effective njLEP, $(\gamma_i - \gamma_e) = \pm 4\Omega_{ie}$, we obtain two degenerate pairs of eigenvalues for the effective Liouvillian superoperator:
\begin{equation}
    \begin{aligned}
        \lambda_{1,2}&=-\frac{\gamma_i+\gamma_e}{2},\\
        \lambda_{3,4}&=-\frac{\gamma_i+\gamma_e}{2}-2\Gamma
    \end{aligned}
\end{equation}

The first degenerate pair corresponds to a second-order EP, while the second pair corresponds to a diabolical point (DP). At a DP, only the eigenenergies become degenerate while the corresponding eigenvectors do not coalesce. This result can be confirmed by computing the Jordan normal form.

As we see, for a specific perturbation, it is possible to split the original EP only partially. In this case, the Jordan blocks are transformed as
\begin{equation}\label{eq:transformation_of_EPs_qubit}
\begin{aligned}
    J_3^{\mathcal{L}^\prime_{\text{eff}}} &\left( -\frac{ \gamma_i + \gamma_e }{2} \right)  
    \rightarrow    \\
    &J_2 \left( -\frac{ \gamma_i + \gamma_e }{2} \right)\oplus  J_1 \left(-\frac{\gamma_i+\gamma_e}{2}-2\Gamma \right), \\
    J_1^{\mathcal{L}^\prime_{\text{eff}}} &\left( - \frac{ \gamma_i + \gamma_e }{2} \right) 
    \rightarrow J_1 \left( -\frac{\gamma_i+\gamma_e}{2}-2\Gamma \right).
\end{aligned}
\end{equation}

\subsection{Non-Hermitian qutrit}\label{sec: examples qutrit}
\subsubsection{Effective njLEPs in a non-Hermitian qutrit}

In case of an effective non-Hermitian qutrit, the situation becomes richer albeit more complicated.
Tuning the effective non-Hermitian Hamiltonian to a third-order EP results in a three-block EP of the effective no-jump Liouvillian. Perturbing this more complicated EP with quantum jump terms in turn leads to a wider range of possibilities.

\begin{figure}[t]
    \centering
    \includegraphics[width=.7\linewidth]{ 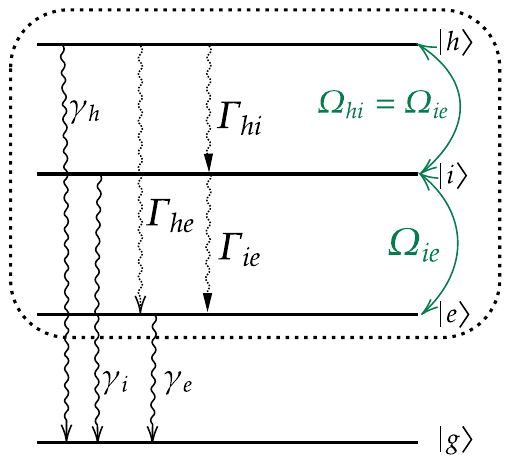}
    \caption{Formation of a non-Hermitian qutrit via a dissipative four-level system.  We consider two equal drives ($\Omega_{hi} = \Omega_{ie}$). For simplicity, we assume that all parameters in the system are real.}
    \label{fig: effective_qutrit_with_decay}
\end{figure}

Similarly to the qubit case, we consider a system with four levels $\ket{h}$, $\ket{i}$, $\ket{e}$ and $\ket{g}$ ordered from highest to lowest energy. In the absence of the dissipative terms connecting the three excited states, the NHH for the qutrit composed of levels $\ket{h}$, $\ket{i}$ and $\ket{e}$ becomes
\begin{equation}
     \hat{H}_{\text{eff}} = \begin{pmatrix}
       \Delta_h - i \frac{{\gamma}_{h}  }{2} & \Omega_{hi} & \Omega_{he} \\
        \Omega_{hi} & \Delta_i - i \frac{{\gamma}_{i}}{2} & \Omega_{ie} \\
        \Omega_{he} & \Omega_{ie} & \Delta_e -i \frac{\gamma_{e}}{2} 
    \end{pmatrix}.
\end{equation}
The terms $\Omega_{hi}$, $\Omega_{he}$ and $\Omega_{ie}$ represent the coupling rates between the respective pairs of states; $\gamma_{h,i,e}$ are the dissipation rates from the levels $\ket{h}$, $\ket{i}$, $\ket{e}$ respectively into the ground state $\ket{g}$. To simplify the discussion, we are going to put $\Delta_{h,i,e}=0$, $\Omega_{he}=0$ and $\Omega_{hi}=\Omega_{ie}$. This simplified model is illustrated in Fig.~\ref{fig: effective_qutrit_with_decay}.

To obtain a triple root, we first make the Hamiltonian traceless:
\begin{equation}
    \hat H_\mathrm{eff}^\prime = \hat H_\mathrm{eff} +i\frac{\gamma_h+\gamma_i+\gamma_e}{6} \hat{\mathbb{1}}_{3\times 3},
\end{equation}
and then look for the parameters, for which the characteristic polynomial simplifies:
\begin{equation}
    \det\qty(\varepsilon \hat{\mathbb{1}} - \hat H_\mathrm{eff}^\prime)\equiv \varepsilon^3.
\end{equation}
The resulting condition of triple degeneracy is
\begin{equation}
    2\gamma_i =\gamma_h+\gamma_e,\qquad \Omega_{ie} = \pm\frac{\gamma_h-\gamma_e}{4\sqrt{2}}\label{eq:triple-degeneracy}
\end{equation}
If $\gamma_h\neq\gamma_e$, this triple root corresponds to a third-order EP of the effective NHH $\hat H_\mathrm{eff}$ with eigenvalue $\varepsilon_\mathrm{EP}=-i\gamma_i/2$.

If $\hat H_\mathrm{eff}$ is tuned to this third-order EP, we deduce from Eq.~\eqref{eq: general relation} that the effective no-jump Liouvillian has a multi-block njLEP with block sizes 5, 3 and 1:
\begin{equation}
    \mathcal{L'}_{\text{eff}} \sim J_5 \qty (\lambda_{\mathcal{L}'_{\mathrm{eff}}}) \oplus J_3 \qty (\lambda_{\mathcal{L}'_{\mathrm{eff}}}) \oplus J_1 \qty (\lambda_{\mathcal{L}'_{\mathrm{eff}}}),
    \label{eq: qutrit Jordan blocks}
\end{equation}
where
\begin{equation}
    \lambda_{\mathcal{L}'_{\mathrm{eff}}} = 2\im{\varepsilon_\mathrm{EP}}  = -\gamma_{i}
\end{equation}

\subsubsection{\label{sec: quantum jumps qutrit}Quantum jumps as perturbations of the system}

To study the effect of the quantum jumps, we include the dissipation within the effective non-Hermitian qutrit.
Similar to the case of the qubit, we consider two different examples of the dissipation:
\begin{enumerate}
\item[(i)] dissipation from level $\ket{h}$ into levels $\ket{i}$ and $\ket{e}$, and from level $\ket{i}$ into level $\ket{e}$. We assume for simplicity that all decay rates for the excited levels are equal, such that $\Gamma_{hi} = \Gamma_{he} = \Gamma_{ie} = \Gamma$. Furthermore, we assume these rates satisfy the condition $\Gamma \ll \gamma_{h}, \gamma_{i}, \gamma_{e}$. The implications of non-identical decay rates are discussed at the end of this section.

\begin{align}
\BL_\mathrm{eff}^\mathrm{(i)} & = \BL_\mathrm{eff}^\prime + \Gamma \qty(D\qty[\ket{i}\bra{h}] + D\qty[\ket{e}\bra{h}] + D\qty[\ket{e}\bra{i}]),\\
D\qty[\hat L] &= \hat L\otimes \hat L^* - \frac{\hat L^\dagger\hat L\otimes\mathbb{1} + \mathbb{1}\otimes\hat L^T\hat L^*}{2},\nonumber
\end{align}
and
\item[(ii)] a particular variant of the bit-phase-flip error that swaps levels $\ket{h}$ and $\ket{e}$ while applying a $\pi$ phase shift to both
\begin{equation}
    \BL_\mathrm{eff}^\mathrm{(ii)} = \BL_\mathrm{eff}^\prime + \Gamma D\qty[\hat L_\mathrm{ph}],\qquad \hat L_\mathrm{ph} = \begin{pmatrix}
         0 & 0 & -1 \\
         0 & 1 & 0 \\
         -1 & 0 & 0
    \end{pmatrix}\label{eq:Leff-qutrit-ii}
\end{equation}
\end{enumerate}

Similarly to the non-Hermitian qubit case, in both cases (i) and (ii), $\Gamma$ is a non-negative coefficient that determines the rates of the non-unitary terms. In case (i), it describes the decay rate for each pair of levels of the non-Hermitian qutrit, whereas in case (ii), it determines the rate of the bit-phase-flip channel. The quantum channels in cases (i) and (ii) are treated as distinct models and are analyzed independently.

Since $\BL_\mathrm{eff}$ is now a $9\times9$ matrix, it is impossible to find its spectrum analytically in general. Instead, we have to treat the quantum jump terms as perturbations.
The characteristic polynomial of $\BL_\mathrm{eff}$ can be written as
\begin{equation}
    \det\qty(\lambda\mathbb{1} - \BL_\mathrm{eff}(\Gamma)) = \lambda^9 + a_1(\Gamma)\lambda^8 + \dotsc+a_8(\Gamma)\lambda + a_9(\Gamma),
\end{equation}
By looking at the leading order terms in the expansions of the characteristic polynomial's coefficients
\begin{equation}
 a_k(\Gamma) = \alpha_k\Gamma^{\beta_k} + o\qty(\Gamma^{\beta_k}),
\end{equation}
one can in turn deduce the leading order terms in the expansion of the polynomial roots $\lambda_i(\Gamma)$.
This constitutes the essence of the Newton diagram technique~\cite{Moro2003}.

\textbullet\ In case (i), we can simplify the problem by noticing that the characteristic polynomial is exactly divisible by a polynomial of the third order. The roots of this factor as well as the corresponding eigenvectors can be determined analytically. The leading in $\Gamma$ behavior for the rest of the eigenvalues can be determined by applying the Newton diagram technique to the quotient resulting from division by this factor (see Appendix~\ref{sec:appD_Newton_diagram} for more details).
The results are summarized as follows: the Jordan block of size $5$ splits into the five branches of a complex $z^{1/5}$ root:
\begin{multline}
    J_5^{\BL_\mathrm{eff}^\prime}(\lambda_{\BL_\mathrm{eff}^\prime}) \to \\
    \lambda_r \approx \lambda_{\BL_\mathrm{eff}^\prime} \!+ \!\qty (\frac{15(\gamma_h-\gamma_e)^4}{512})^{1/5} \Gamma^{1/5} e^{i \frac{2 \pi r }{5} },\quad  r = 1\dots.
\end{multline}
The block of size $3$ also splits completely, albeit with non-generic exponents. The corresponding eigenvalues come from the factor of the third order and hence are known exactly:
\begin{equation}
\begin{aligned}
    J_3^{\BL_\mathrm{eff}^\prime}(\lambda_{\BL_\mathrm{eff}^\prime}) &\to \\
    &\lambda_{6}= \lambda_{\BL_\mathrm{eff}^\prime}-\Gamma\\
    &\lambda_{7,8} = \lambda_{\BL_\mathrm{eff}^\prime} - \Gamma \pm \frac{1}{2} \sqrt{\Gamma \qty (\Gamma + \gamma_h  - \gamma_e )}
    \end{aligned}
\end{equation}
Finally, the block of size $1$ shifts linearly in eigenvalue,
\begin{equation}
    J_1^{\BL_\mathrm{eff}^\prime}(\lambda_{\BL_\mathrm{eff}^\prime}) \to J_1(\lambda_{\BL_\mathrm{eff}^\prime} - 16\Gamma/15),\\
\end{equation}

\textbullet\ In case (ii), the perturbation matrix is highly symmetric and we can factorize the characteristic polynomial into polynomials of order not larger than $3$. As a consequence, all the eigenvalues and eigenvectors can be computed analytically. Hence, the initial Jordan blocks are transformed as follows:
\begin{align}\label{eq: transformation of EPs qutrit}
    &J_5^{\mathcal{L'}_{\text{eff}} } \qty (\lambda_{\BL_\mathrm{eff}^\prime})  \to J_3 \qty (\lambda_{\BL_\mathrm{eff}^\prime}) \oplus J_2 \qty (\lambda_{\BL_\mathrm{eff}^\prime} - 2 \Gamma), \\
     &J_3^{\mathcal{L'}_{\text{eff}} } \qty (\lambda_{\BL_\mathrm{eff}^\prime})  \to J_1\qty (\lambda_{\BL_\mathrm{eff}^\prime} ) \oplus J_2 \qty (\lambda_{\BL_\mathrm{eff}^\prime} - 2 \Gamma),\\
     &J_1^{\mathcal{L'}_{\text{eff}} } \qty (\lambda_{\BL_\mathrm{eff}^\prime})  \to  J_1 \qty (\lambda_{\BL_\mathrm{eff}^\prime} ).
\end{align}

As we have demonstrated, both the type of the njLEP and the structure of the dissipative terms determine the behavior of the Liouvillian spectrum in the vicinity of the njLEP.

For both the non-Hermitian qubit and qutrit, in case (i), the degeneracy of the njLEP is completely lifted by a small perturbation $\Gamma$. For the qubit, the blocks split in agreement with their sizes, as expected for a generic perturbation~\cite{lidskii_perturbation_1966,moro_lidskii--vishik--lyusternik_1997}: the block of size $3$ splits as $\propto \Gamma^{1/3}$, while the eigenvalue of the block of size $1$ shifts linearly. For the qutrit, this is only true for the blocks of sizes $5$ and $1$, which split as $\propto \Gamma^{1/5}$ and shift linearly, respectively. This happens because the choice $\Gamma_{hi} = \Gamma_{he} = \Gamma_{ie} = \Gamma$ preserves certain symmetries, making the perturbation non-generic. In the case of a general perturbation, where the decay rates are distinct but have fixed ratios, the block of size $3$ for the qutrit would also split as $\propto \Gamma^{1/3}$, where $\Gamma$ now represents a common prefactor~\cite{GS_SS}.

In case (ii), for both the non-Hermitian qubit and qutrit, the perturbation is too symmetric, so that the njLEP structure is not entirely destroyed by the addition of quantum jump terms.

\section{Dynamical evolution of qubit and qutrit level populations near EPs}\label{sec:numerical results}

\begin{figure}[h!]
  \centering
  \begin{minipage}[h!]{0.45\textwidth}
    \includegraphics[width=\textwidth]{ 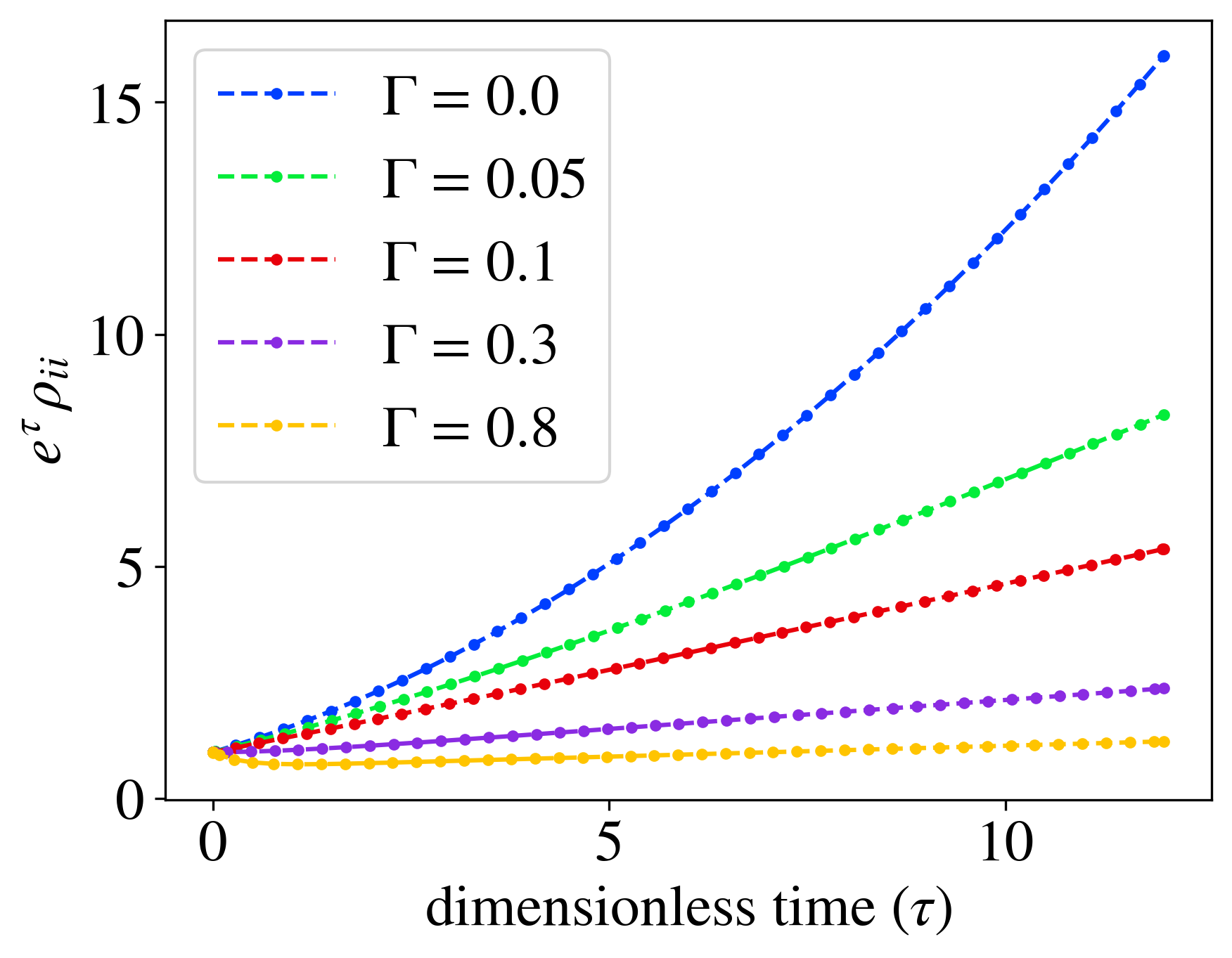}
    \caption{Evolution of the population of the  \(\ket{i}\)-level of the effective qubit, factoring out the main exponential decay. The time dependence of \(e^{\tau} \rho_{ii}(\tau)\) is shown for the parameters \(\Omega_{ie} = \frac{\gamma_{e} - \gamma_{i}}{4} \), \(\gamma_{e} = 0.9\), \(\gamma_{i} = 0.2\), and for various values of \(\Gamma\). Here, $\tau = t(\gamma_i+\gamma_e)/2$.}
  \label{fig: rho_ii evolution qubit}
  \end{minipage}
  \hfill
  \begin{minipage}[h!]{0.45\textwidth}
    \includegraphics[width=\textwidth]{ 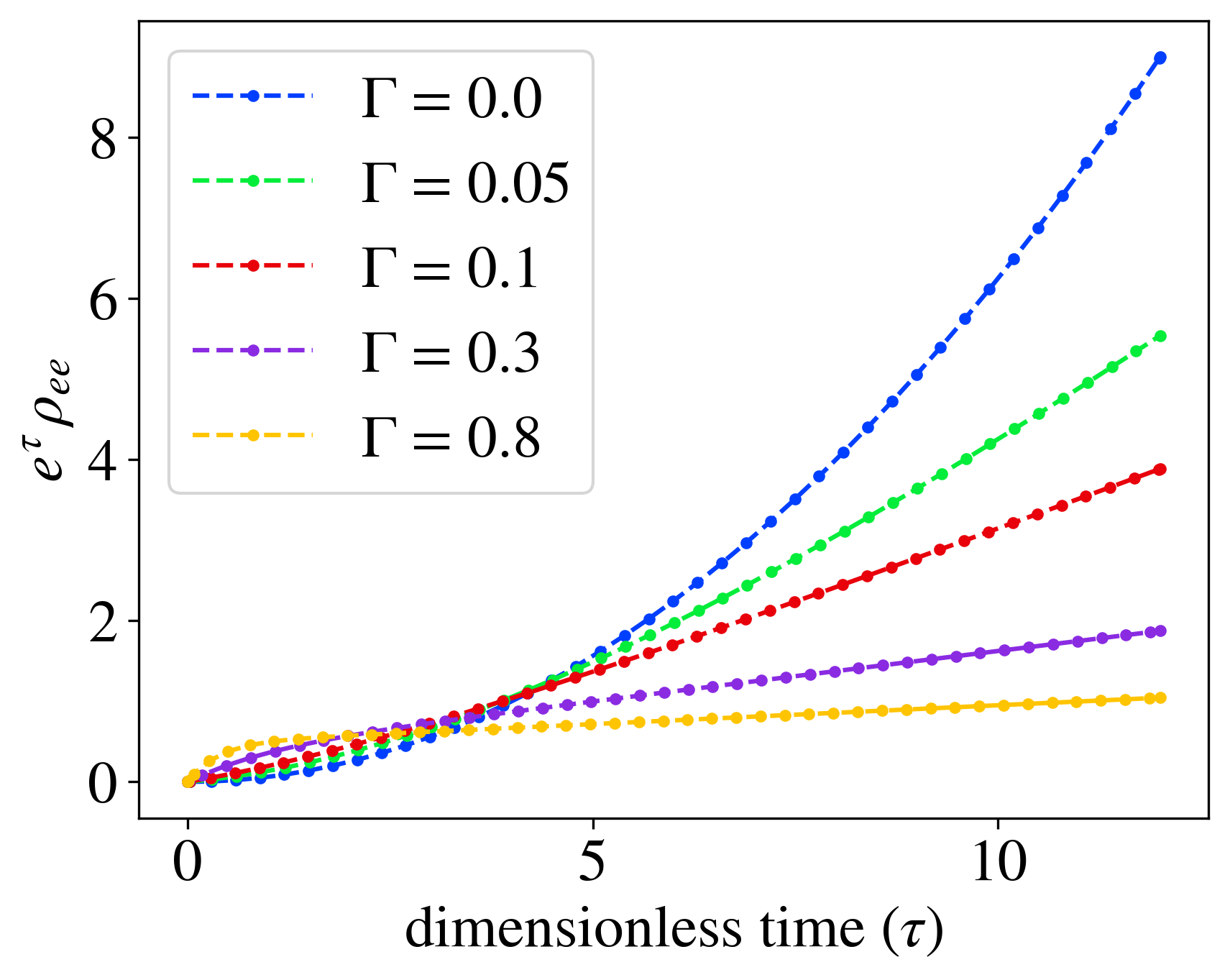}
    \caption{Evolution of the population of the  \(\ket{e}\)-level of the effective qubit, factoring out the main exponential decay. The time dependence of \(e^{\tau} \rho_{ee}(\tau)\) is shown for the parameters \(\Omega_{ie} = \frac{\gamma_{e} - \gamma_{i}}{4}\), \(\gamma_{e} = 0.9\), \(\gamma_{i} = 0.2\), and for various values of \(\Gamma\). Here, $\tau = t(\gamma_i+\gamma_e)/2$.}
  \label{fig: rho_ee evolution qubit}
  \end{minipage}
\end{figure}

\begin{figure}[h!]
  \centering
  \begin{minipage}[h!]{0.45\textwidth}
    \includegraphics[width=\textwidth]{ 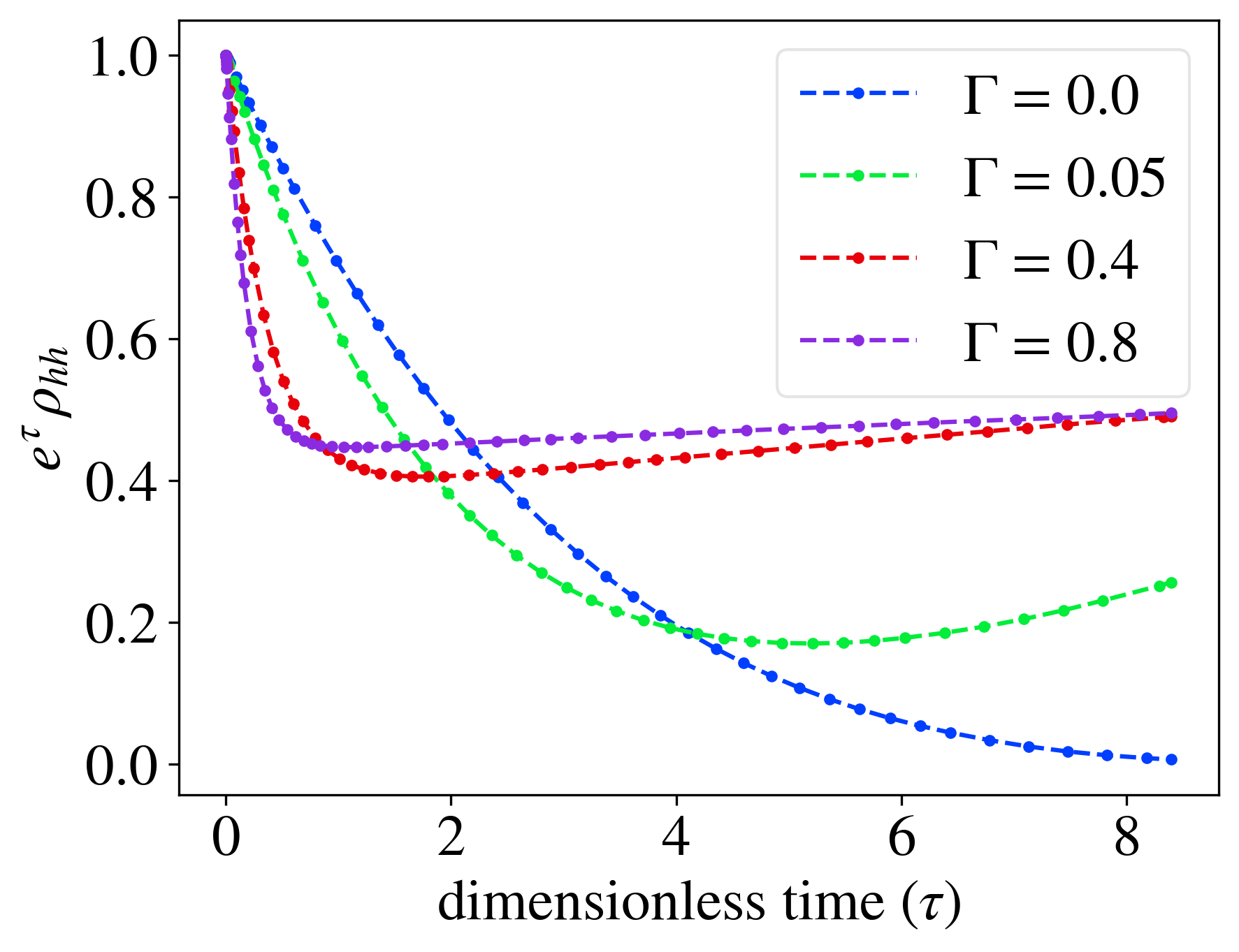}
    \caption{Evolution of the population of the \(\ket{h}\)-level of the effective qutrit, factoring out the main exponential decay. The time dependence of \(\rho_{hh}(\tau)\) is shown for the parameters \(\Omega_{ie} = \frac{\gamma_{h} - \gamma_{e}}{4\sqrt{2}} \), $\gamma_i = (\gamma_h+\gamma_e)/2$, \(\gamma_{h} = 0.4\), \(\gamma_{e} = 0.2\), and for various values of \(\Gamma\). Here, $\tau=t\gamma_i$.}
    \label{fig:rho_hh qutrit}
  \end{minipage}
\end{figure}

\begin{figure}[h!]
  \centering
  \begin{minipage}[h!]{0.45\textwidth}
    \includegraphics[width=\textwidth]{ 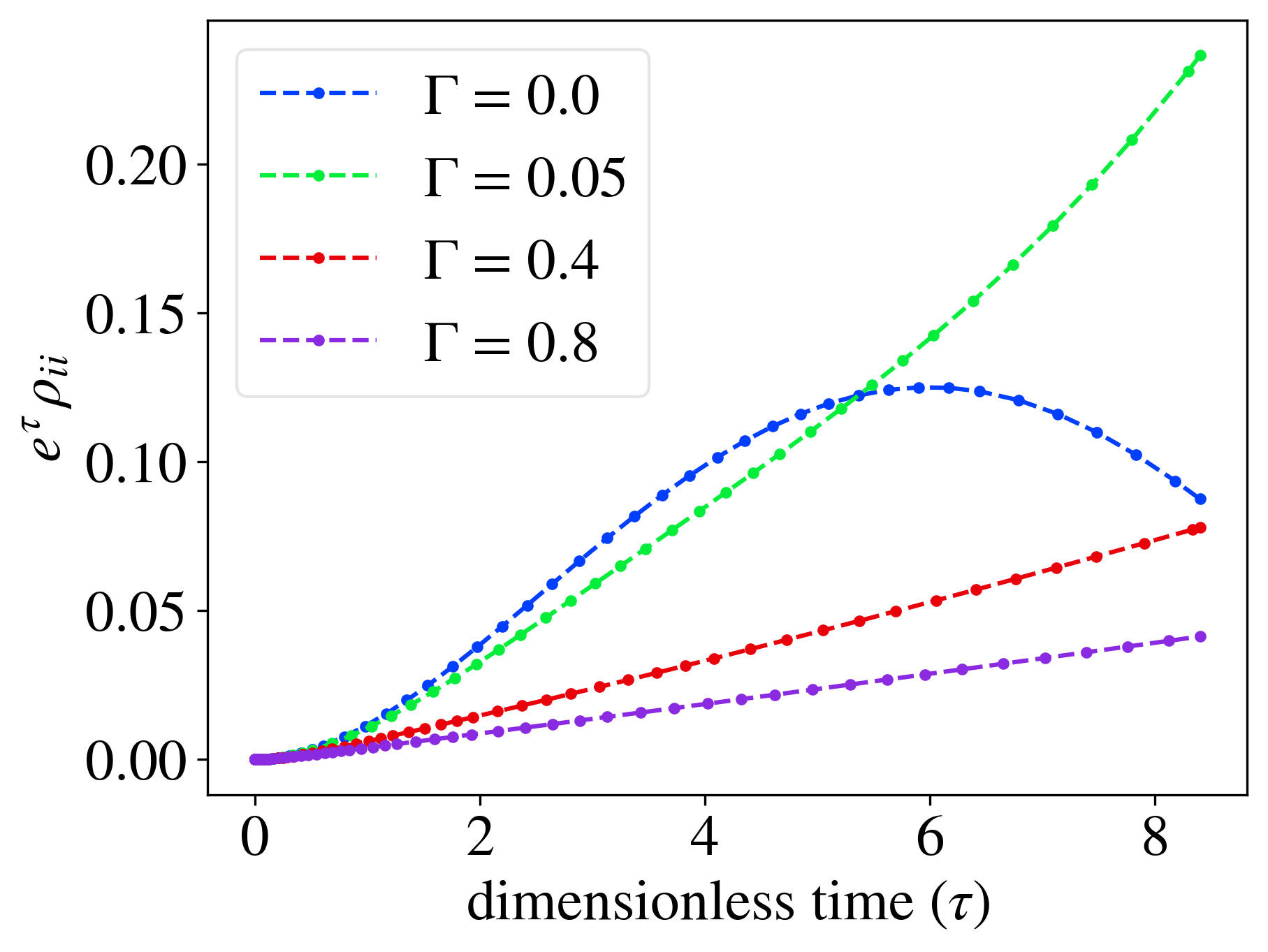}
   \caption{Evolution of the population of the \(\ket{i}\)-level of the effective qutrit, factoring out the main exponential decay. The time dependence of \(\rho_{ii}(\tau)\) is shown for the parameters \(\Omega_{ie} = \frac{\gamma_{h} - \gamma_{e}}{4\sqrt{2}} \), $\gamma_i = (\gamma_h+\gamma_e)/2$, \(\gamma_{h} = 0.4\), \(\gamma_{e} = 0.2\), and for various values of \(\Gamma\). Here, $\tau=t\gamma_i$.}
   \label{fig:rho_ii qutrit}
  \end{minipage}
    \hfill
  \begin{minipage}[h!]{0.45\textwidth}
    \includegraphics[width=\textwidth]{ 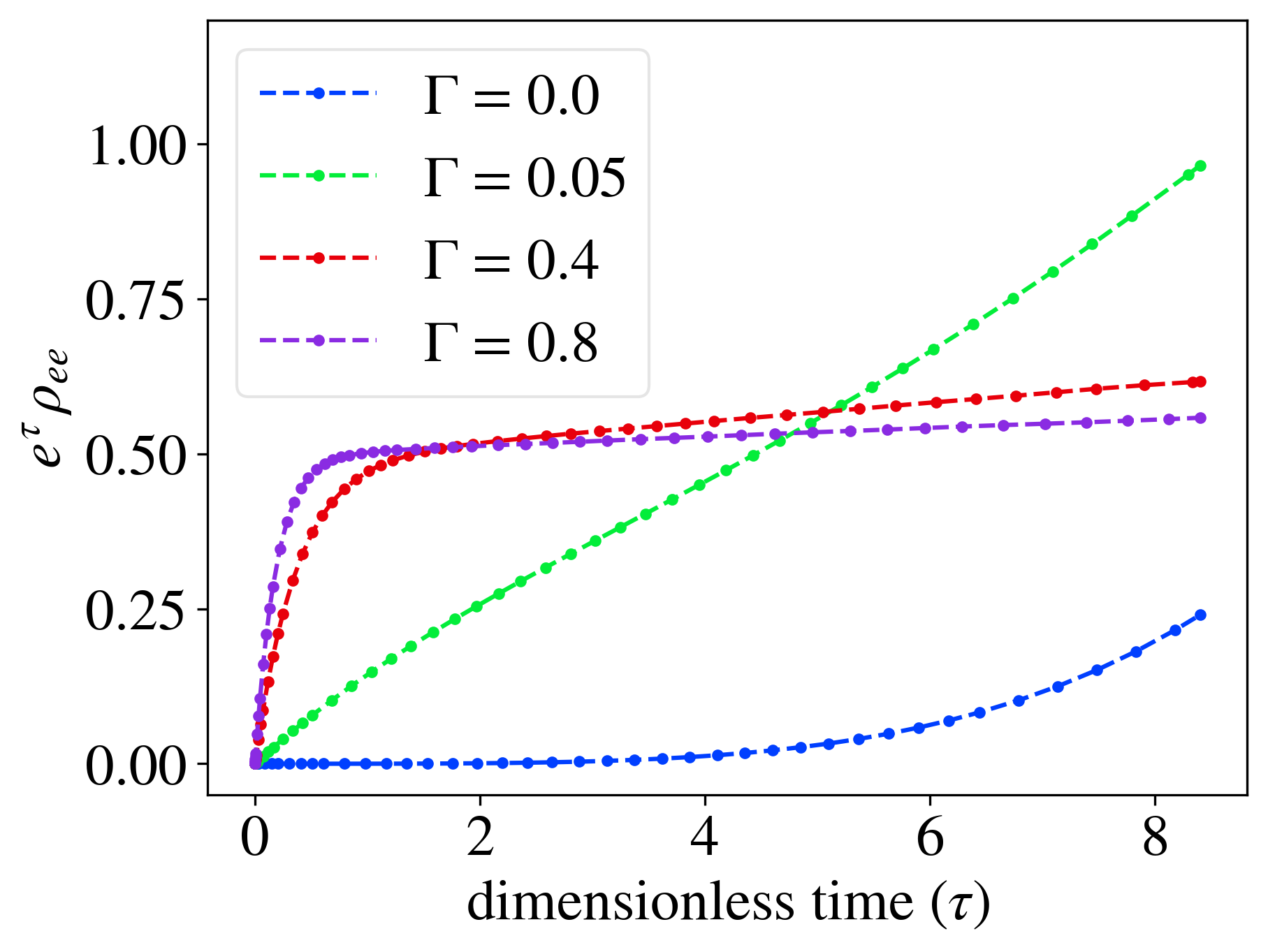}
    \caption{Evolution of the population of the \(\ket{e}\)-level of the effective qutrit, factoring out the main exponential decay. The time dependence of \(\rho_{ee}(\tau)\) is shown for the parameters \(\Omega_{ie} = \frac{\gamma_{h} - \gamma_{e}}{4\sqrt{2}} \), $\gamma_i = (\gamma_h+\gamma_e)/2$, \(\gamma_{h} = 0.4\), \(\gamma_{e} = 0.2\), and for various values of \(\Gamma\). Here, $\tau=t\gamma_i$.}
    \label{fig:rho_ee qutrit}
  \end{minipage}
\end{figure}

Given the peculiar multi-block character of the njLEPs, it is crucial to look for their signatures, both without the quantum jump terms and in their presence. One of such possibilities is to look at the population dynamics of the excited states.

The time-dependence of the projected density matrix is determined by the exponential of the effective Liouvillian superoperator $e^{\BL_\mathrm{eff}t}$ (see Eqs.~\eqref{eq:rho-proj-dynamics} and~\eqref{eq:proj-exponential}). The form of the exponential is easily determined in the Jordan basis of $\BL_\mathrm{eff}$ using the known property that
\begin{equation}
e^{J_n(\lambda)t} = e^{\lambda t}
\begin{pmatrix}
        1 & t & \frac{t^2}{2} & \frac{t^3}{3!} & \dotsb & \frac{t^{n-1}}{(n-1)!}\\
        0 & 1 & t & \frac{t^2}{2} &  \dotsb & \frac{t^{n-2}}{(n-2)!}\\
        0 & 0 & 1 & t & \dotsb &  \frac{t^{n-3}}{(n-3)!}\\
        \vdots & \ddots & \ddots & \ddots & \ddots & \vdots\\
        0&\dotsb& 0& 0& 1&t\\
        0 & \dotsb & 0 & 0 & 0& 1 \\
    \end{pmatrix}
\end{equation}
As we see, in the presence of EPs, the usual pure exponential decay (eigenvalues of $\BL_\mathrm{eff}$ have negative real part) acquires polynomial factors in $t$. For general conditions, the order of the polynomial is $(n-1)$, where $n$ is the size of the largest Jordan block. This implies that the higher the order of the EP (the size of the largest block), the slower is the evolution to the steady state.

We begin with the case of the qubit with combined bit- and phase-flip error (case (ii) in Sec.~\ref{sec: quantum jumps}).
We assume that the effective no-jump Liouvillian is tuned to the multi-block EP 
\[
\Omega_{ie} = \pm \frac{\qty(\gamma_{e} - \gamma_{i})}{4}
\]
and vary $\Gamma$ from zero to finite positive values. This way the initial $(3,1)$-block EP splits into an EP2 and a DP at different eigenvalues. As the result, the order of the polynomial prefactor is lowered from $2$ to $1$.

This behavior can be observed in Figs.~\ref{fig: rho_ii evolution qubit} and \ref{fig: rho_ee evolution qubit}. Here, we employ the initial condition $\rho_0 = \begin{pmatrix}
    1 & 0 \\
    0 & 0
\end{pmatrix}$. To focus on the polynomial prefactor, we remove the dominant exponential decay factor $e^{-t \frac{\qty(\gamma_{i} + \gamma_{e})}{2}} = e^{-\tau}$, associated with the least-decaying mode of the effective Liouvillian superoperator. As it can be observed, population dynamics at short times is very sensitive to the value of $\Gamma$ for $\Gamma$ close to zero, especially for the $\ket{e}$-level.

Next, we perform the same analysis for the qutrit and the perturbation in the form of the dephasing error (case (ii) in Sec.~\ref{sec: quantum jumps qutrit}). This way, we expect the order of the polynomial prefactors to lower from $4$ to $2$. We choose the parameters to tune the effective no-jump Liouvillian to the EP described by the multi-block structure in Eq.~\eqref{eq: qutrit Jordan blocks}.  To do so, we impose the conditions~\eqref{eq:triple-degeneracy} and use the projected density matrix
\begin{equation}
   \rho_0 = \begin{pmatrix}
        1 & 0 & 0 \\
        0 & 0 & 0 \\
        0 & 0 & 0
    \end{pmatrix}\label{eq:qutrit-init-rho}
\end{equation}
as the initial value.
Similar to the case of the qubit, we remove the exponential decay factor $e^{-\tau}=e^{-\gamma_i t}$, associated with the least-decaying eigenmode of the effective no-jump Liouvillian, and observe the strong sensitivity of the short-time dynamics to the value of $\Gamma$ for $\Gamma$ close to zero.

Interestingly, the curves (see Figs.~\ref{fig:rho_hh qutrit}--\ref{fig:rho_ee qutrit}) exhibit linear behavior at larger values of $\Gamma$ and $\tau$. This is due to the fact that, for the initial condition~\eqref{eq:qutrit-init-rho}, the coefficient of $t^2$ in the polynomial prefactor is very small.

To summarize the results of this section, the time dependence of observables acquires polynomial prefactors in the presence of LEPs, whose order is determined by the size of the largest Jordan block. Perturbing LEPs lifts the degeneracy and decreases the block sizes, which is reflected in the time dependence of observables at small $\Gamma$.

\section{Detecting LEPs using the quantum geometric tensor}\label{sec:QGT}

Another interesting possibility to probe the structure of EPs in our systems is to study the quantum geometric tensor (QGT), proposed in Ref.~\cite{provost_riemannian_1980} for Hermitian (closed) quantum systems, and recently extended to non-Hermitian systems~\cite{Solnyshkov_D_PhysRevB.103.125302,Orlov_P_PhysRevB.111.L081105}. There, it was shown to diverge in the vicinity of  EPs~\cite{Qing_Liao_PhysRevLett.127.107402}.

We follow the definition of the QGT originally introduced in Ref.~\cite{brody_information_2013} and later generalized in Ref.~\cite{hu_generalized_2024, orlov_adiabatic_2025}. In our case, however, we apply it not to the eigenvectors of $\hat H_\mathrm{eff}$, but to the vectorized eigenmatrices of the superoperator $\BL_\mathrm{eff}$.
\begin{equation}
        Q_{n, \mu \nu}^{LR} =  \left( \partial_\mu \vec{\rho}_n^{\, L} \right)^{\dagger} \cdot \partial_\nu \vec{\rho}_n^{\, R} 
        -\left( \left( \partial_\mu \vec{\rho}_n^{\, L} \right)^{\dagger} \cdot \vec{\rho}_n^{\, R} \right)  
         \left( \left ( \vec{\rho}_n^{\, L} \right)^{\dagger} \cdot \partial_\nu \vec{\rho}_n^{\, R} \right).
\end{equation}
Here, $\vec{\rho}_n^{\, R}$ and $\vec{\rho}_n^{\, L}$ are the right and left eigenmatrices in vectorized form respectively 
\begin{equation}
\mathcal{L}_{\text{eff}}{\vec{\rho}}_n^{\, R} = \lambda_n \vec{\rho}_n^{\, R}, \quad \mathcal{L}_{\text{eff}}^\dagger \vec{\rho}_n^{\, L} = \lambda_n^*\vec{\rho}_n^{\, L},
\end{equation}
that satisfy the biorthonormality condition $\left(\vec{\rho}_m^{\, L}\right)^{\dagger} \cdot  \vec{\rho}_n^{\, R}  = \delta_{mn}$.
We assume $\BL_\mathrm{eff}$ to depend on the set of physical parameters $\boldsymbol q=q^\mu$, so $\partial_\mu=\partial/\partial q^\mu$ denote the derivatives with respect to this set. Note that thus defined $Q_{n, \mu \nu}^{LR}$ is gauge-invariant with respect to the transformations
\begin{equation}
    \vec{\rho}_n^{\, R}  \rightarrow e^{\chi_n(\boldsymbol{q})} \vec{\rho}_n^{\, R} , \quad 
    \vec{\rho}_n^{\, L}  \rightarrow e^{ -\chi_n^*(\boldsymbol{q}) } \vec{\rho}_n^{\, L} ,
\end{equation}
preserving the biorthonormality condition. Here, $\chi_n(\boldsymbol q)$ are arbitrary differentiable complex-valued functions.


\begin{widetext}
\vspace*{-2 em}
\begin{figure}[h!]
    \centering
    {\includegraphics[width=\textwidth]{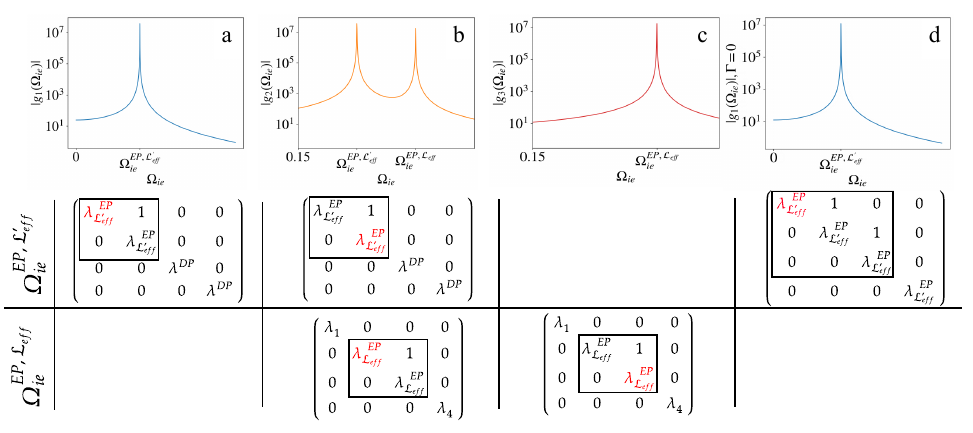}}
    \caption{
    Quantum metric of an effective qubit with phase and bit-flip errors.  
    (a)–(c) Dependence of the quantum metric on $\Omega_{ie}$ for parameters $\gamma_{e} = 0.9$, $\gamma_{i} = 0.1$, and $\Gamma = 0.3$;  
    (d) Dependence of the quantum metric on $\Omega_{ie}$ for parameters $\gamma_{e} = 0.9$, $\gamma_{i} = 0.1$, and $\Gamma = 0.0$.  
    The divergence of the quantum metric indicates the presence of EPs.  
    Below the plots, we display the Jordan normal form of $\BL_\mathrm{eff}$ at the special values of $\Omega_{ie}$ where the quantum metric diverges.  
    The levels for which the quantum metric is computed are marked in red. The $y$-axis is displayed on a $\log_{10}$ scale.}
    \label{fig:QGT qubit}
\end{figure}

\end{widetext}

\begin{figure}[h!]
            \centering
            \includegraphics[scale=.35]{ 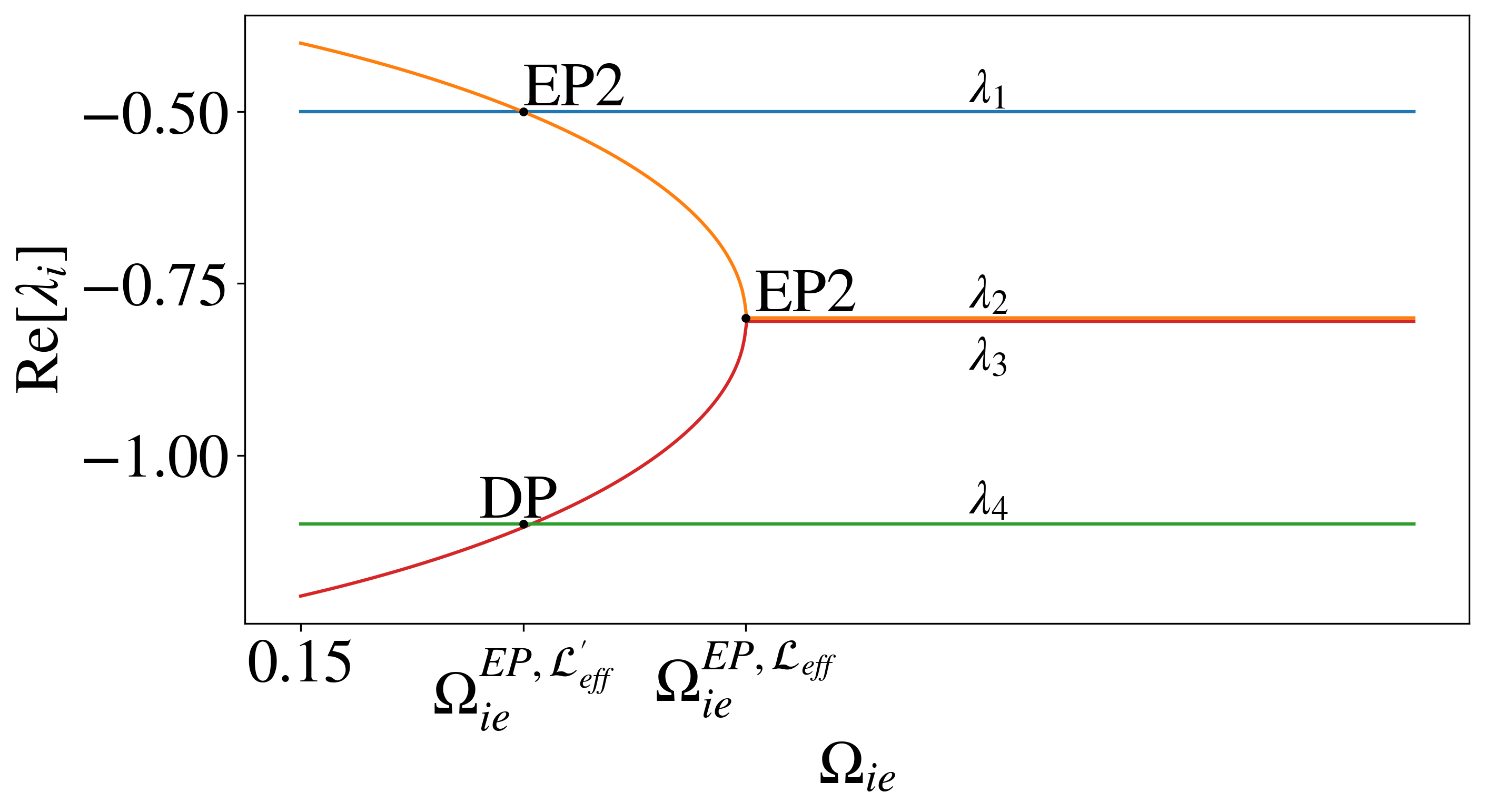}
            \includegraphics[scale=.35]{ 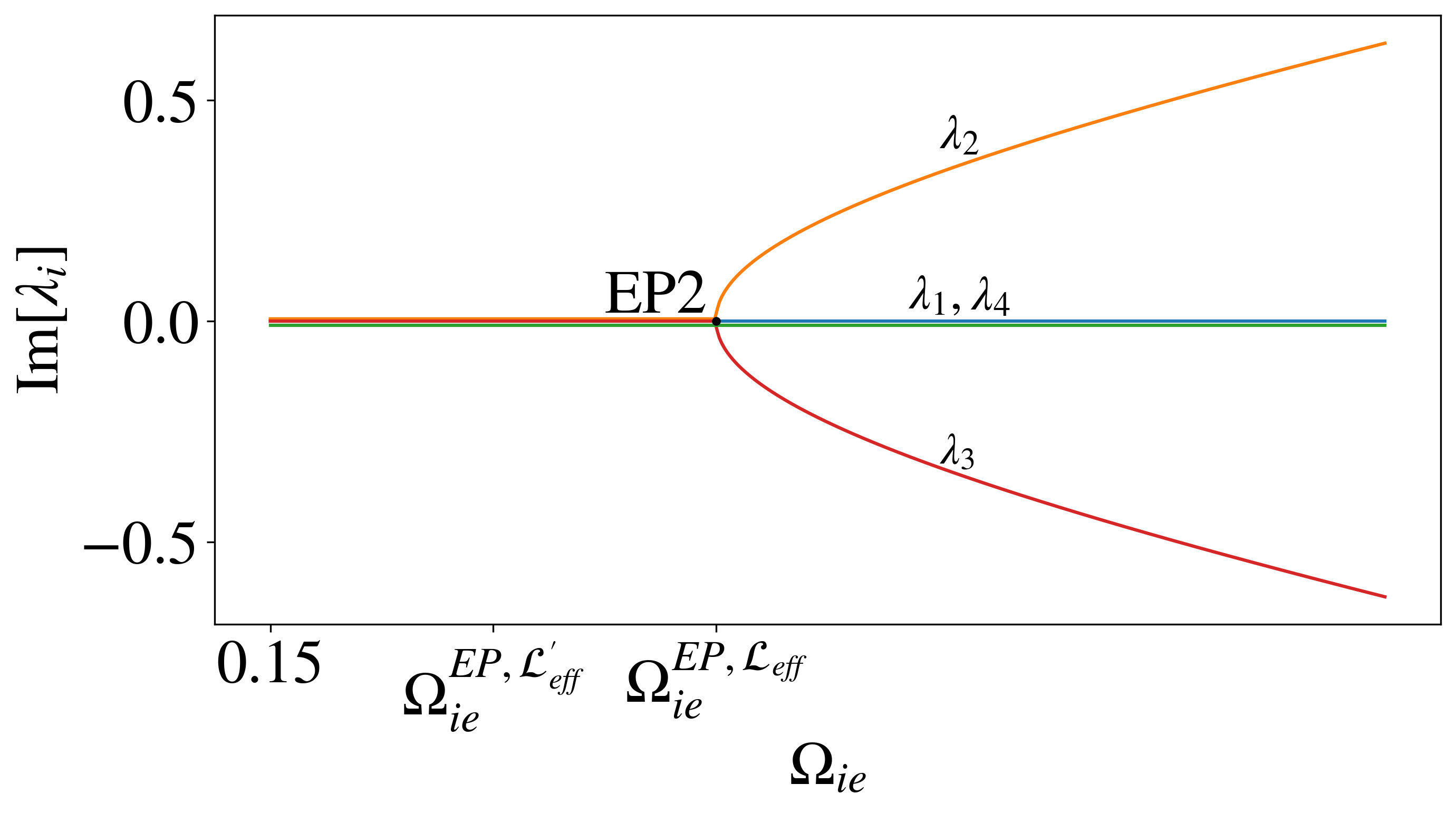}
            \caption{
            Real and imaginary parts of the spectrum of an effective qubit with phase- and bit-flip errors. The figure shows the dependence of the eigenspectrum of the effective non-Hermitian qubit (Eq.~\ref{eq:eigenvalues qubit}) on $\Omega_{ie}$, with parameters $\gamma_{e} = 0.9$, $\gamma_{i} = 0.1$, and $\Gamma = 0.3$. EPs, their orders, and DPs are marked on the graph. The curves have been slightly offset for clarity.}
            \label{fig: qubit_eigenvalues_spectrum}
\end{figure}

For concreteness, we keep $\Gamma$ fixed and only consider the dependence of the QGT on $q^1 = \Omega_{ie}$. We then focus our attention on the quantum metric --- the real part of the QGT tensor:
\begin{equation}
    g_{n}(\Omega_{ie}) = \re{\qty[Q^{LR}_{n,11}]}.
\end{equation}

In Fig.~\ref{fig:QGT qubit}, we present the QGT for the effective qubit subject to the combined phase- and bit-flip error (case (ii) of Sec.~\ref{sec: quantum jumps}). Only the levels that form the Jordan block of size 3 at $\Gamma=0$ are shown. Correspondingly, in Fig.~\ref{fig: qubit_eigenvalues_spectrum}, we provide the real and imaginary parts of the eigenvalues as the functions of $\Omega_{ie}$. The line colors label the eigenvalues and are consistent between the two figures.

The QGT diverges at two distinct values of $\Omega_{ie}$: $\Omega_{ie}^{EP,\BL^\prime_\mathrm{eff}}=(\gamma_i-\gamma_e)/4$, which corresponds to the multi-block EP at zero $\Gamma$, and $\Omega_{ie}^{EP,\BL_\mathrm{eff}}=\sqrt{(\gamma_i-\gamma_e)^2+4\Gamma^2}/4$. This picture is typical for a perturbed EP3~\cite{starkov_formation_2023}: at non-zero $\Gamma$, the Jordan block of size $3$ is split into two EP2-s. Moreover, we observe that only one level is involved in both EP2-s.

It is interesting to note that at $\Omega_{ie} = \Omega_{ie}^{EP,\BL^\prime_\mathrm{eff}}$ no square-root splitting of the imaginary parts of the eigenvalues appears, which is typical of a EP2 (compare it with the behavior of the imaginary parts at $\Omega_{ie}^{EP,\BL_\mathrm{eff}}$). This has a symmetry reason: We have mentioned in the Introduction that the effective Liouvillian superoperator has a generalized $\BP\BT$-symmetry. As $\Omega_{ie}$ is varied, the system touches the boundary of the $\BP\BT$-unbroken phase at this point without crossing into the $\BP\BT$-broken phase.

\begin{figure}[ht] 
  \label{ fig7} 
  \begin{minipage}[b]{0.5\linewidth}
    \centering
    \includegraphics[width=\linewidth]{ 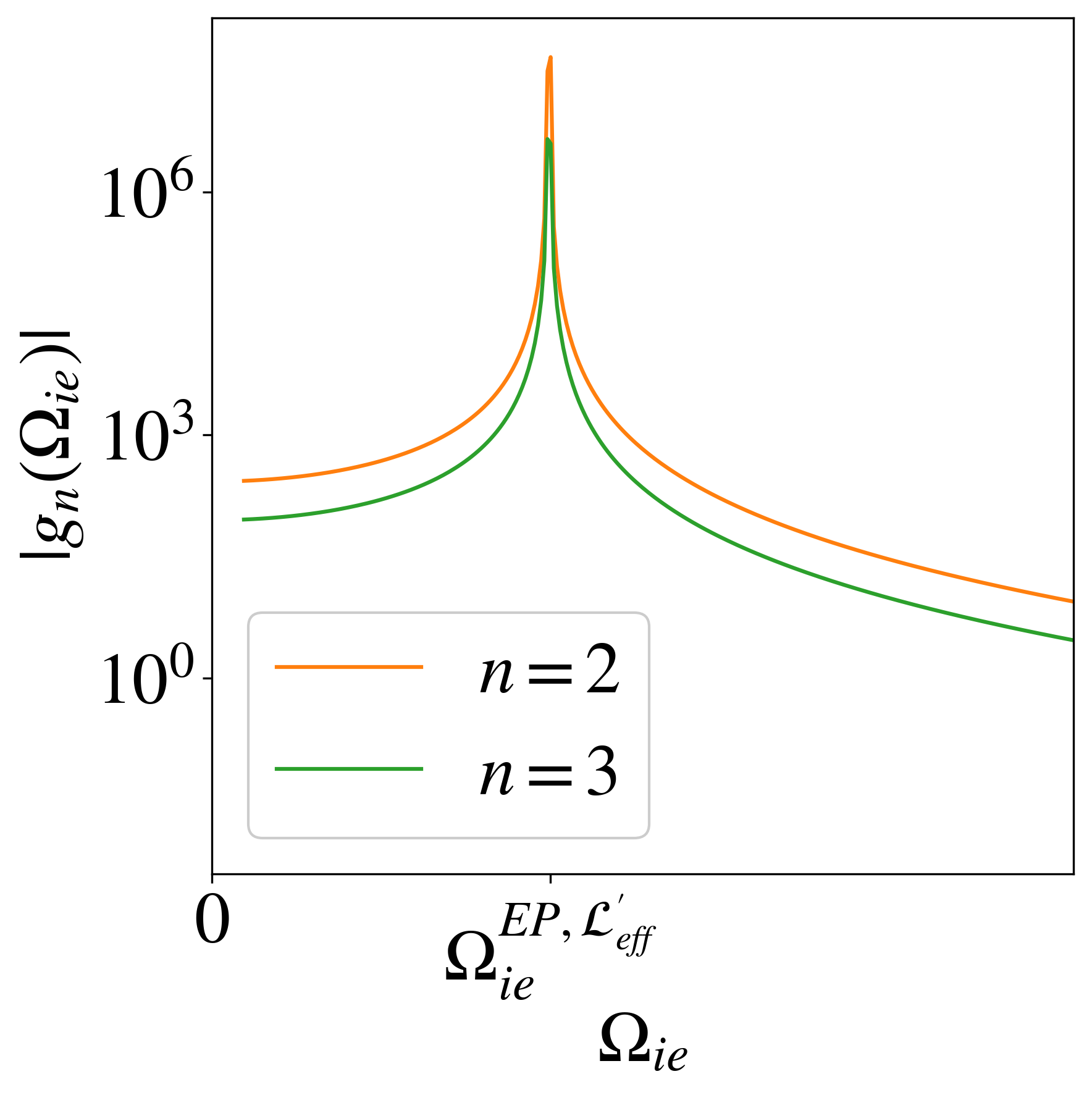} 
  \end{minipage}
  \begin{minipage}[b]{0.5\linewidth}
    \centering
    \includegraphics[width=\linewidth]{ 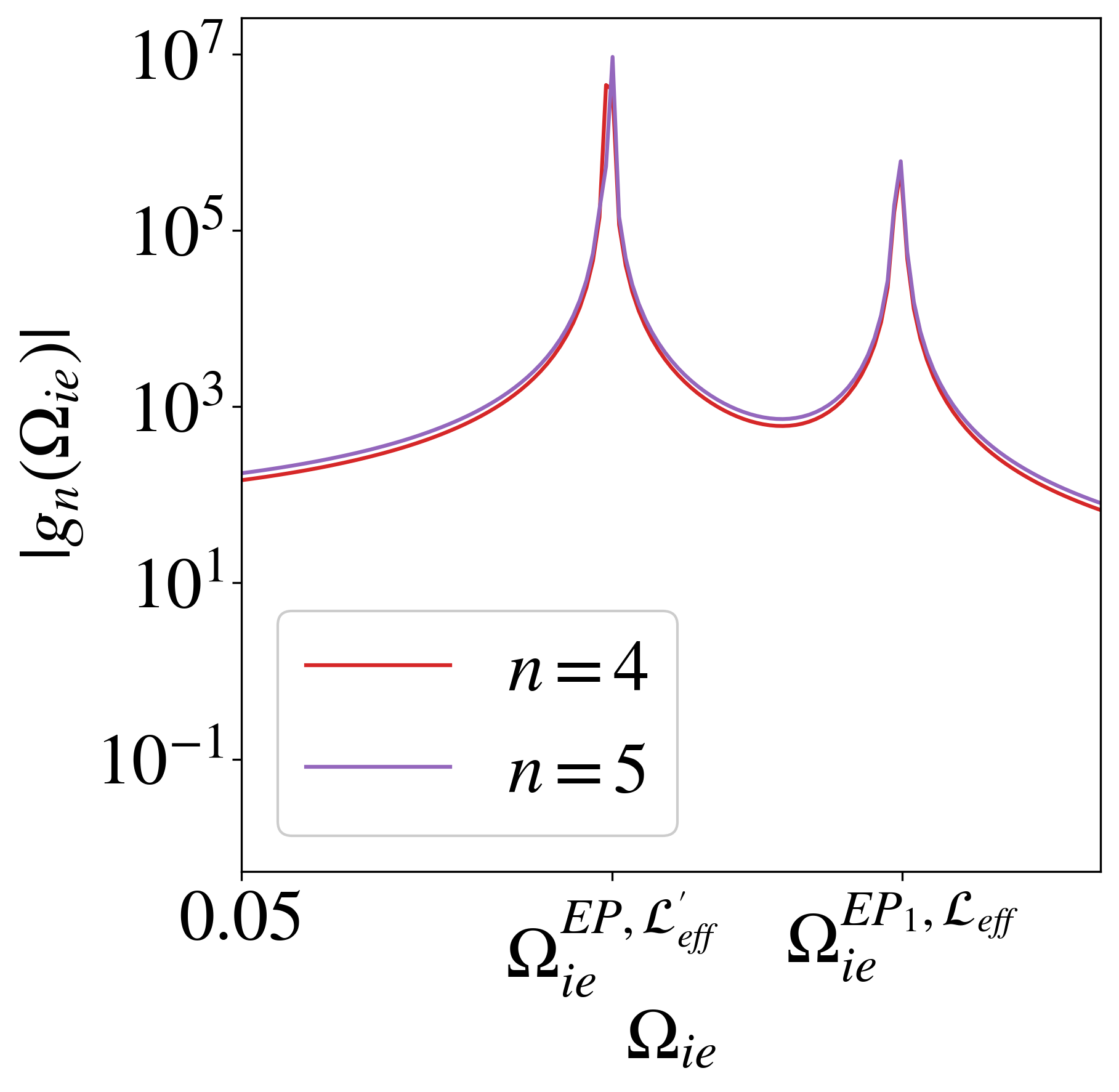} 
  \end{minipage} 
   \begin{minipage}[b]{0.5\linewidth}
    \centering
    \includegraphics[width=\linewidth]{ 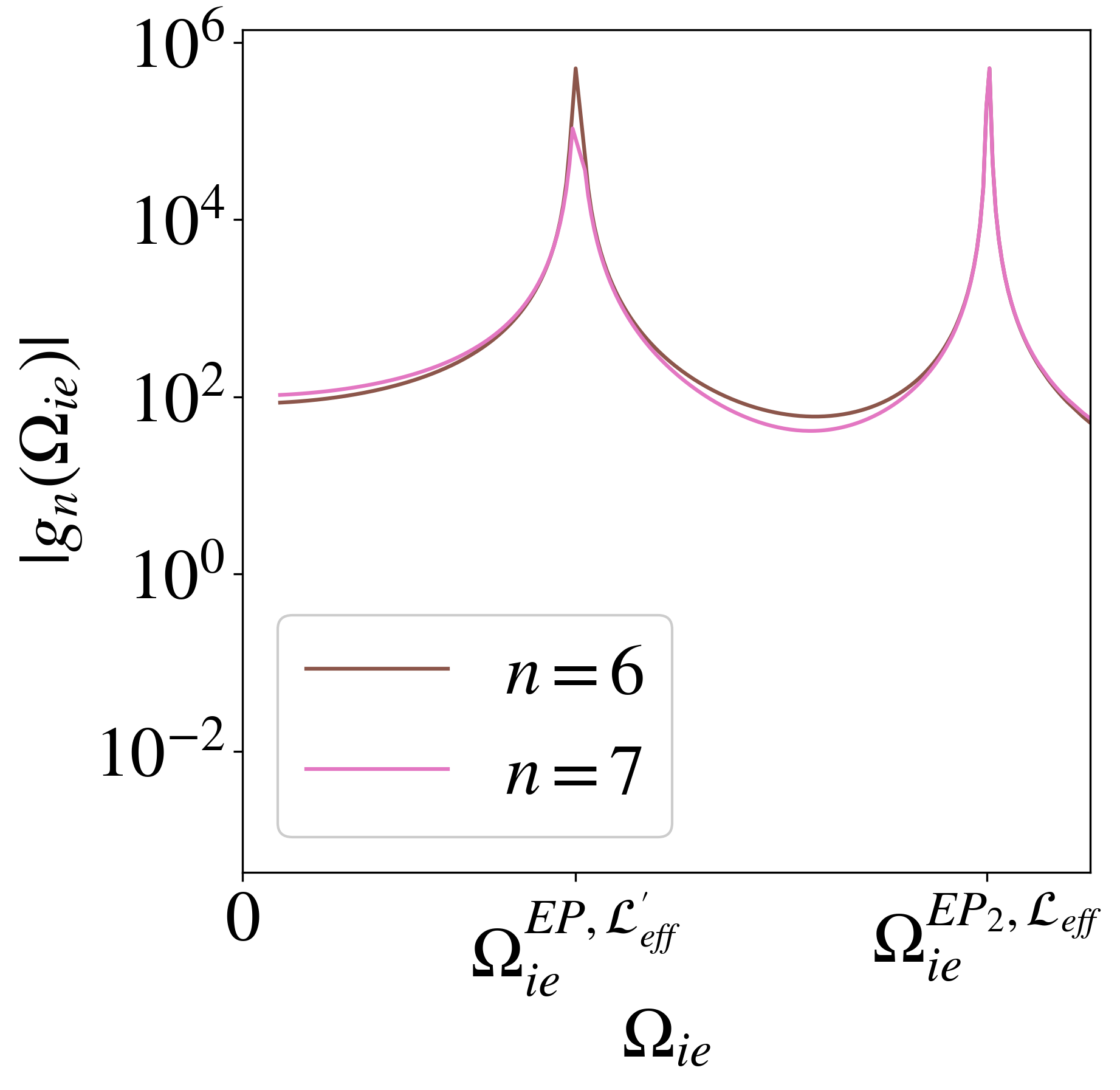} 
  \end{minipage}
  \begin{minipage}[b]{0.5\linewidth}
    \centering
    \includegraphics[width=\linewidth]{ 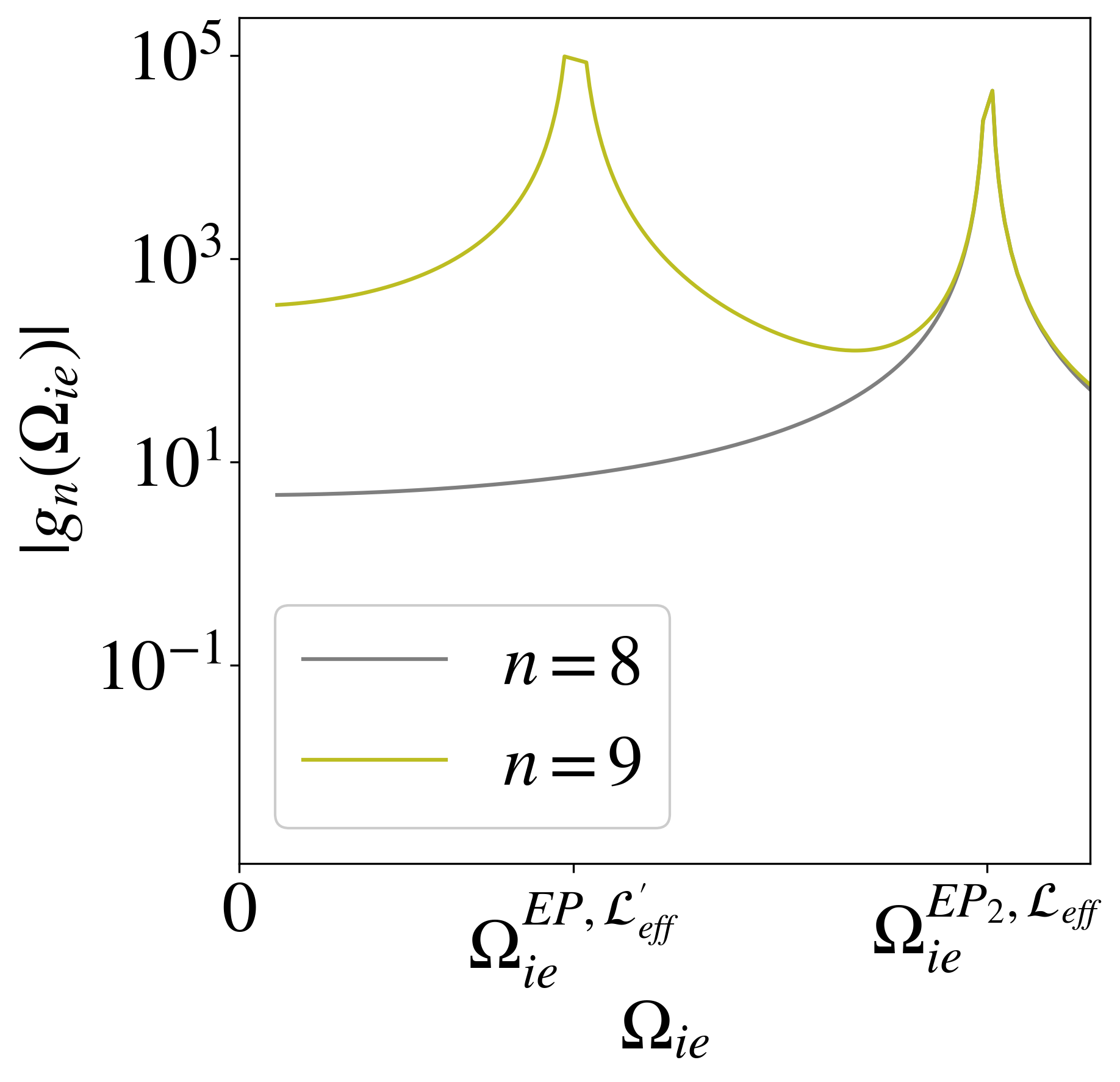} 
  \end{minipage} 
  \caption{
    Quantum metric of an effective qutrit with phase error.  
    Quantum metric dependence on $\Omega_{ie}$ with parameters $\gamma_{h} = 0.8$, $\gamma_e = 0.2$, $\gamma_{i} = (\gamma_{h} + \gamma_{e}) / 2$, and $\Gamma = 0.3$.  The divergence of the quantum metric indicates the presence of EPs. The $y$-axis is displayed on a $\log_{10}$ scale.}
    \label{fig:QGT qutrit}
\end{figure}

 \begin{figure}[ht] 
            \centering
            \includegraphics[scale=.31]{ 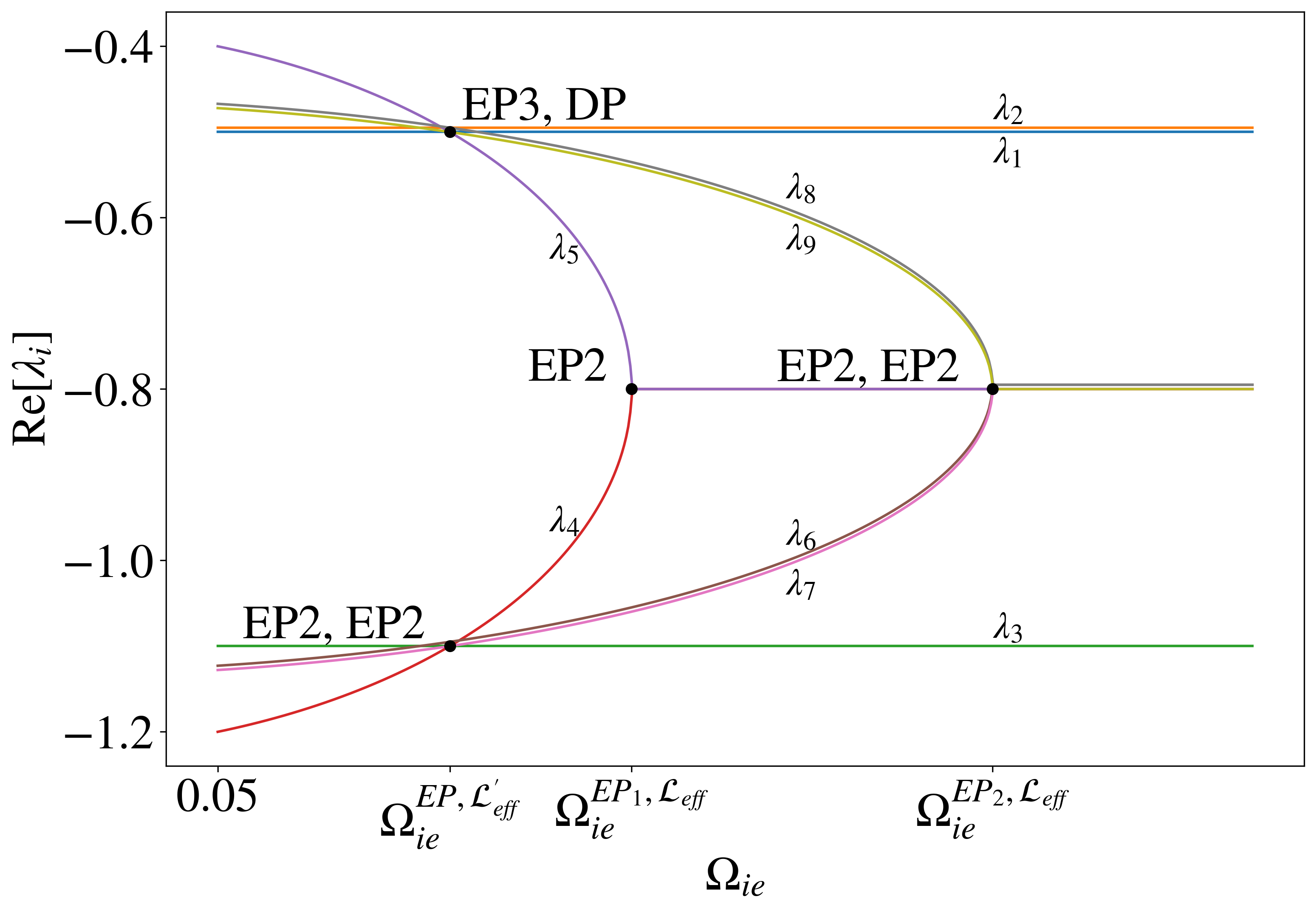}
            \includegraphics[scale=.31]{ 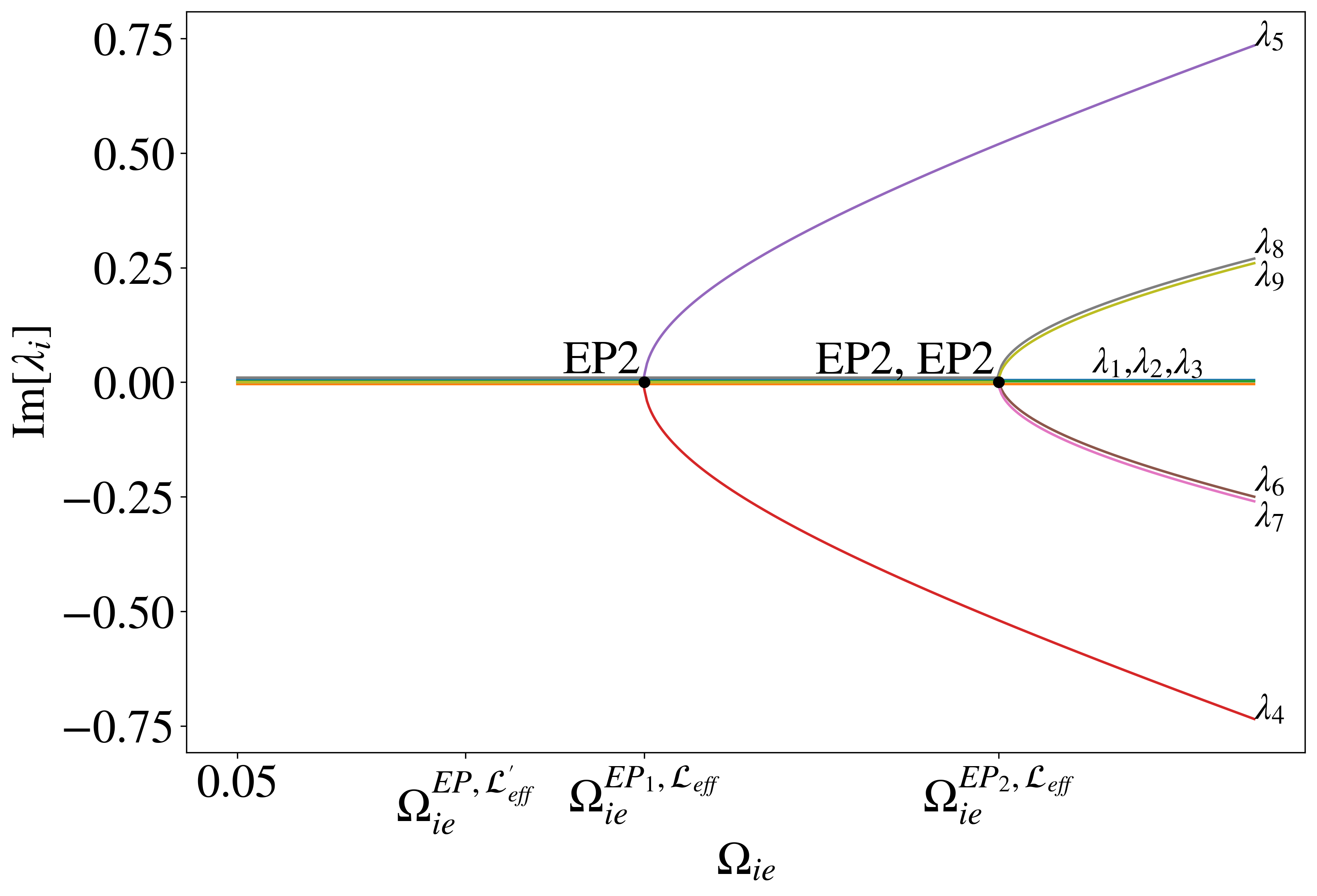}
            \caption{
            Real and imaginary parts of the spectrum of an effective qutrit with phase error. The figure shows the dependence of the eigenspectrum of the effective non-Hermitian qubit (Eq.~\ref{eq:eigenvalues qubit}) on $\Omega_{ie}$, with parameters $\gamma_{h} = 0.8$, $\gamma_e=0.2$ $\gamma_{i} = (\gamma_h+\gamma_e)/2$, $\Gamma = 0.3$. EPs, their orders, and DPs are marked on the graph. The curves have been slightly offset for clarity.}
            \label{fig:qutrit_eigenvalues_spectrum}
    \end{figure}

Similar results are obtained when analyzing the QGT of an effective qutrit in presence of a dephasing error (case (ii) of Sec.~\ref{sec: quantum jumps qutrit}). We display the QGT and the eigenvalue curves respectively in Figs.~\ref{fig:QGT qutrit} and~\ref{fig:qutrit_eigenvalues_spectrum}. When $\Gamma = 0$, we detect only a single value of $\Omega_{ie}$ that leads to an EP, denoted as $\Omega_{ie}^{\text{EP}, \mathcal{L'}_{\text{eff}}}$. However, when $\Gamma > 0$, three critical values of $\Omega_{ie}$ are observed, one of which coincides with $\Omega_{ie}^{\text{EP}, \mathcal{L'}_{\text{eff}}}$. The analytical expressions for the eignevalues and the critical values of $\Omega_{ie}$ are listed in Appendix~\ref{sec:appE_eigenvals_qubit}.

The structure in the effective qutrit case is considerably more intricate due to the presence of multiple Jordan blocks in the no-jump Liouvillian, which are partially split by the quantum jump term. The overall structure of the QGT in this case matches the theoretical predictions in Eq.~(\ref{eq: transformation of EPs qutrit}). When the drive is tuned to $\Omega_{ie} = \Omega_{ie}^{\text{EP}, \mathcal{L'}_{\text{eff}}}$, the eigenvalues $\lambda_2$, $\lambda_5$, and $\lambda_9$ coalesce to form an EP3, corresponding to the Jordan block $J_3\qty(-\gamma_{i})$. The eigenvalues $\lambda_{3,4}$ and their associated eigenvectors form an EP2 described by the Jordan block $J_2\qty(-2\Gamma - \gamma_{i})$. A similar structure is observed for $\lambda_{6,7}$ and their eigenvectors, which also form an EP2 of the same type, $J_2\qty(-2\Gamma - \gamma_{i})$.

Analyzing the cases of the effective qubit and qutrit, we conclude that the QGT can indeed serve as a useful tool to study the structure of the LEPs: Looking at the QGT for different levels, we can identify which of the levels are involved in the formation of an EP.

\section{Conclusions and outlook} \label{sec:con}

In this work, we explore structure and behavior of the exceptional points in open quantum systems governed by Lindbladian dynamics. We establish the general correspondence between HEPs and njLEPs. As we have demonstrated, no-jump Liouvillians naturally exhibit a novel type of EPs, characterized by multiple Jordan blocks at the same eigenvalue, which we dub multi-block EPs. This is crucial to properly analyze the spectrum of the Liouvillian superoperator, even when the quantum jump terms are taken into account.

We illustrate our general findings based on effective qubit and qutrit models derived from dissipative three- and four-level systems, where the ground state serves as a sink.
In these two models, we focus on the observables sensitive to the structure of EPs: population dynamics of the excited states, and the extension of the Quantum Geometric Tensor to the Liouvillian superoperators.
Population dynamics is sensitive to the order of EPs through the polynomial corrections to the exponential decay. The divergence of the QGT computed for different levels allows us to identify EPs and the levels involved in their formation.

The most intriguing feature of multi-block EPs is that the blocks could, in principle, be merged together even with an arbitrarily small perturbation. A generic perturbation of an EP with detuning $\epsilon$ leads to the eigenenergy splitting that scales as $\propto \epsilon^{1/m}$, where $m$ is the size of the largest Jordan block. Merging the blocks can increase $m$, thus making the system's response to small perturbations amplified, which is advantageous for EP-based quantum sensors~\cite{Wiersig:20sensors, Hodaei2017}. In addition to that, increasing the Jordan block size results in a slower decay towards the steady state due to the power-law prefactors (see Sec.~\ref{sec:numerical results}). Theoretically, it may help prolong the lifetime of excited states, which could be useful for quantum computing.


Mathematically speaking, we can demonstrate this point by considering a multi-block EP of the effective qutrit from Sec.~\ref{sec: examples qutrit}. The $t$-perturbed Jordan block structure can be written as
\begin{equation}\label{eq:merging-the-blocks-example}
    \begin{pmatrix}
        \lambda & 1 & 0 & 0 & 0 & 0 & 0 & 0 & 0\\
        0 & \lambda & 1 & 0 & 0 & 0 & 0 & 0 & 0\\
        0 & 0 & \lambda & 1 & 0 & 0 & 0 & 0 & 0\\
        0 & 0 & 0 & \lambda & 1 & 0 & 0 & 0 & 0\\
        0 & 0 & 0 & 0 & \lambda & \textcolor{red}{t} & 0 & 0 & 0\\
        0 & 0 & 0 & 0 & 0 & \lambda & 1 & 0 & 0\\
        0 & 0 & 0 & 0 & 0 & 0 & \lambda & 1 & 0\\
        0 & 0 & 0 & 0 & 0 & 0 & 0 & \lambda & 0\\
        0 & 0 & 0 & 0 & 0 & 0 & 0 & 0 & \lambda\\
    \end{pmatrix}\sim J_8(\lambda)\oplus J_1(\lambda)
\end{equation}
For $t=0$, we get $\BL^\prime_\mathrm{eff}$ in its Jordan basis, as shown in Eq.~\eqref{eq: qutrit Jordan blocks}. However, for any non-zero $t$, the Jordan blocks of sizes $5$ and $3$ are merged to form a single block of size $8$.

Physically, engineering specifically this form of perturbation might not even be feasible within the standard Liouvillian framework\footnote{A ``reverse engineering'' approach reveals that this perturbation requires terms proportional to $\left(\ket{i} \bra{h}\right) \otimes \left(\ket{e} \bra{h} \right) + \left(\ket{e} \bra{h}\right) \otimes \left(\ket{i}\bra{h}\right)$.}. To realize block-merging in practice, we need to characterize all the perturbations that result in block-merging and then identify those that are consistent with the structure of the quantum jump terms in the Liouvillian superoperator (see Eq.~\eqref{eq:eff_Lindbladian}). This is a challenging task that goes beyond the scope of the present paper.

\begin{acknowledgments}
    This work was supported by the DFG-SFB 1170 (Project-ID: 258499086) and EXC2147 ctd.qmat (Project-ID: 390858490).
\end{acknowledgments}

\appendix

\section{Connection between njLEPs and HEPs} \label{sec: app A EPs Lindblad and NHH general relation}

In this section, we follow the approach used in Ref.~\cite{TRAMPUS1966242} to construct the Jordan basis for the Kronecker sum of the $m\times m$ matrix $A$ and the $n\times n$ matrix $B$
\begin{equation}
    \mathscr{L} = A\otimes \mathbb{1}_n + \mathbb{1}_m\otimes B.
\end{equation}
Eventually, we are going to substitute $A=-i\hat H_\mathrm{eff}$ and $B=i\hat H^*_\mathrm{eff}$.

Let us for simplicity assume that both $A$ and $B$ host EPs of maximal size:
\begin{equation}
    A\sim J_m(\lambda_a),\qquad B\sim J_n(\lambda_b).
\end{equation}
Let $v_i$ and $w_j$ for $i=1,2,\dots,m$ and $j=1,2,\dots,n$ be the Jordan basis of the matrices $A$ and $B$ respectively, then
\begin{equation}
    (A - \lambda_a\mathbb{1}_m)v_i =v_{i-1},\qquad (A - \lambda_a\mathbb{1}_m)v_1 = 0,
\end{equation}
\begin{equation}
    (B - \lambda_a\mathbb{1}_m)w_i =w_{i-1},\qquad (A - \lambda_a\mathbb{1}_m)w_1 = 0.
\end{equation}
As we see, $v_1$ and $w_1$ are the eigenvectors of $A$ and $B$ respectively. Following Ref.~\cite{TRAMPUS1966242}, we call $v_i$ and $w_i$ generalized eigenvectors of grade $i$, which comes from
\begin{equation}
    (A-\lambda_a\mathbb{1}_m)^i v_i =0, \qquad (B-\lambda_a\mathbb{1}_m)^i w_i =0.
\end{equation}
In the following, it is convenient to introduce
\begin{equation}
    A_- = A-\lambda_a\mathbb{1}_m, \quad B_- = B - \lambda_b\mathbb{1}_n,\quad \mathscr{L}_- = \mathscr{L}-\lambda\mathbb{1}_{mn},
\end{equation}
which can be viewed as analogues of lowering operators. The Jordan basis of $\mathscr{L}$ is constructed in a manner that is reminiscent of the way the basis of the total angular momentum is built when the two angular momenta are added in Quantum Mechanics. The length of the Jordan chain plays the role of the absolute value of the momentum, while the grade of the vector --- the role of the 
projection.

The set of tensor products $v_i\otimes w_j$ builds the basis for the Kronecker sum $\mathscr{L}$. It is straightforward to check, that $v_1\otimes w_1$ is an eigenvector of $\mathscr{L}$ corresponding to the eigenvalue $\lambda=(\lambda_a+\lambda_b)$. As it turns out, for arbitrary $i$ and $j$, $v_i\otimes w_j$ is a generalized eigenvector of grade $i+j-1$ corresponding to the same eigenvalue. To demonstrate it, we first note that
\begin{equation}
    \mathscr{L}_- = A_-\otimes \mathbb{1}_n + \mathbb{1}_m\otimes B_-
\end{equation}
and that the two terms in the Kronecker sum commute with each other. As a result,
\begin{multline}
    \mathscr{L}_-^k v_i\otimes w_j = \qty[\sum_{l=0}^{k}
    \begin{pmatrix}k \\ l\end{pmatrix}
    A_-^{l}\otimes B_-^{k-l}]v_i\otimes w_j=\\
    \sum_{l=0}^{k} \begin{pmatrix}k \\ l\end{pmatrix} \theta(i-l)\theta(j+l-k) v_{i-l}\otimes w_{j+l-k}.
\end{multline}
Here, zeroth power of the matrix is interpreted as the identity matrix, $\theta(l)$ is the Heaviside theta-function satisfying $\theta(0)=0$. It is straightforward to check that
\begin{equation}
    \mathscr{L}_-^{i+j-2} v_i\otimes w_j = \begin{pmatrix}i+j-2 \\ i-1\end{pmatrix} v_1\otimes w_1\label{eq:last-in-chain}
\end{equation}
and $\mathscr{L}_-^{i+j-1} v_i\otimes w_j=0$.
As shown below, the Jordan basis of $\mathscr{L}$ is formed by linear combinations of $v_i\otimes w_j$ with the same grade.

There is a unique basis vector of the highest grade $\vec{\rho}_{m+n-1,1} = v_m\otimes v_n$. Acting on it repeatedly with $\mathscr{L}_-$, we get the first Jordan chain
\begin{equation}
    \vec{\rho}_{k,1} = \mathscr{L}_-^{m+n-k} \vec{\rho}_{m+n-1,1},\quad k = 1,2,\dots,m+n-1
\end{equation}
where $\vec{\rho}_{1,1}\propto v_1\otimes w_1$, which follows from Eq.~\eqref{eq:last-in-chain}. For the next-to-highest grade $i+j-2$, there are two vectors $v_m\otimes w_{n-1}$ and $v_{m-1}\otimes w_n$, from which we can build two linear independent combinations. One of the combinations is $\vec{\rho}_{m+n-2,1} = v_m\otimes w_{n-1} + v_{m-1}\otimes w_n$ and belongs to the first Jordan chain we construct above. As a result, the second linear combination must belong to the next Jordan chain.
In the general case, ${\vec{\rho}^{\, \prime}} = \alpha v_m\otimes w_{n-1} + \beta v_{m-1}\otimes w_n$ is also a generalized eigenvector of grade $i+j-2$, and $\mathscr{L}_-^{i+j-3}{\vec{\rho}^{\, \prime}}\propto \vec{\rho}_{1,1}$. Different Jordan chains must end with different eigenvectors. The only way to satisfy this condition is to require that $\mathscr{L}_-^{i+j-3}{\vec{\rho}^{\,  \prime}}=0$, {\it i.e.\ }, ${\vec{\rho}^{\, \prime}}$ is of grade $i+j-3$. This gives
\begin{equation}
    \vec{\rho}_{m+n-3,2} = (m-1)v_m\otimes w_{n-1} - (n-1) v_{m-1}\otimes w_n
\end{equation}
and the second Jordan chain
\begin{equation}
    \vec{\rho}_{k,2} = \mathscr{L}_-^{k_2+1-k} \vec{\rho}_{k_2,2}, \quad k = 1,2,\dots, k_2 = m+n-3.
\end{equation}
It is straightforward to check that $\vec{\rho}_{1,2}\neq0$ and is a linear combination of the tensor products of grade $2$. This is no surprise, since there is only one tensor product of grade $1$ which is $\propto \vec{\rho}_{1,1}$.

We can then repeat these steps to find all the Jordan chains. We get by induction that the length of the next Jordan chain is always reduced by $2$. Let us say we construct first $(u-1)$ Jordan chains. At this point, we exhaust all tensor products of grades $1,2,\dots,u-1$ and $m+n-1,m+n-2,\dots,m+n+1-u$. The $u$-th Jordan chain has to start with the linear combination of the tensor products of grade $m+n-u$ and end at the eigenvector composed of the tensor products of grade $u$. The starting vector
\begin{equation}
    \vec{\rho}_{k_u,u} = \sum_{i=0}^{u-1} \alpha_i v_{m-i}\otimes w_{n-u+1+i},\quad k_u = m+n+1-2u
\end{equation}
can then be determined from the condition $\mathscr{L}_-^{k_u}\vec{\rho}_{k_u,u}=0$. The result is ~\cite{TRAMPUS1966242} 
\begin{align}\label{eq: Jordan chain for the Liouvillian superoperator}
\vec{\rho}_{k_u,u} &= \sum_{i=0}^{u-1}(-1)^i \binom{u-1}{i} \frac{(m-1-i)!}{(m-u)!}\times\\ & \frac{(n-u+i)!}{(n-u)!} 
v_{m-i} \otimes w_{n-u+1+i},\nonumber\\
\vec{\rho}_{k,u} &= \mathscr{L}_-^{k_u+1 - k} \vec{\rho}_{k_u,u}, \quad k=1,2,\dots,k_u.
\end{align}
The number of the Jordan chains $\mu$ can be obtained from the condition $\sum_{u=1}^{\mu} k_u = mn$ giving $\mu=\min{(m,n)}$. The resulting Jordan normal form is
\begin{equation}
    \mathscr{L} \sim \bigoplus_{u=1}^{\min{(m,n)}} J_{m+n-(2u-1)}.~\label{eq:single-pair}
\end{equation}

If $A$ and (or) $B$ has multiple Jordan blocks, let us denote the basis vectors as $v_{i,i_c}$ and $w_{j,j_c}$, where $i_c$ and $j_c$ now label the Jordan chains, while $i$ and $j$ denote the grades of the generalized eigenvectors within the chains. Fixing $i_c$ and $j_c$, we can construct the Jordan chains of $\mathscr{L}$ using the procedure we have just described. To obtain the full Jordan basis of $\mathscr{L}$, we need to combine the basis vectors resulting from all $(i_c,j_c)$ pairs. Correspondingly, if we denote by $m_{i_c}$ and $n_{j_c}$ the sizes of the Jordan chains of the matrices $A$ and $B$, we find the Jordan normal form of $\mathscr{L}$ by adding the blocks~\eqref{eq:single-pair} resulting from all $(i_c,j_c)$ combinations:
\begin{equation}
    \mathscr{L} \sim \bigoplus_{i_c,j_c} \bigoplus_{u=1}^{\min{(m_{i_c},n_{j_c})}} J_{m_{i_c}+n_{j_c}-(2u-1)}.\label{eq:all-pairs}
\end{equation}

To apply this result to our case, we need to relate the Jordan basis of $\hat H_\mathrm{eff}$ to that of $-i\hat H_\mathrm{eff}$.
Let's say, the Jordan basis of $\hat H_\mathrm{eff}$ is $h_{l,l_c}$, then the Jordan basis of $-i\hat H_\mathrm{eff}$ is given by $i^{\, L}h_{l,l_c}$,. The Jordan basis for $i\hat H_\mathrm{eff}^*$ is obtained by complex conjugation. After these substitutions, Eq.~\eqref{eq:all-pairs} transforms into Eq.~\eqref{eq:no-jump-jordan-form} of the main text.

\section{Newton diagram in case (ii) of the Section~\ref{sec: quantum jumps qutrit}\label{sec:appD_Newton_diagram}}

Let us compute the characteristic polynomial of $\BL_\mathrm{eff}^\mathrm{(ii)}$ (see Eq.~\eqref{eq:Leff-qutrit-ii}). To simplify the expression, we perform the shift $\lambda \to \tilde{\lambda} = \lambda + \gamma_i$, where $-\gamma_i$ corresponds to the EP of the effective no-jump Liouvillian $\mathcal{L}'_{\mathrm{eff}}$.
The characteristic polynomial then reads (we denote $\tilde{\gamma} = (\gamma_h - \gamma_e)/2$):
\begin{widetext}
\begin{equation}
\begin{aligned}
    \chi(\tilde{\lambda}) &= \frac{1}{32}\qty (\Gamma + \tilde{\lambda}) \qty ( \tilde{\lambda}^2 + 
    2 \Gamma \tilde{\lambda} -\frac{\tilde{\gamma} \Gamma}{2} + \frac{3 \Gamma^2}{4} )
      \\
      &\quad \cdot \bigg{(}-16 \tilde{\gamma}^3 \Gamma^2 (\Gamma + \tilde{\lambda}) + 
  8 \tilde{\lambda} (\Gamma + \tilde{\lambda})^2 (2 \Gamma + \
\tilde{\lambda}) (\Gamma + 2 \tilde{\lambda}) (3 \Gamma + 
     2 \tilde{\lambda}) - \tilde{\gamma}^4 \Gamma (16 \Gamma + 
     15 \tilde{\lambda}) \\
     & \quad \quad + 
  8 \tilde{\gamma}^2 \Gamma (\Gamma + \tilde{\lambda}) (\
\Gamma^2 - 2 \tilde{\lambda}^2) - 
  16 \tilde{\gamma} \Gamma (\Gamma + \tilde{\lambda})^2 (3 \
\Gamma^2 + 10 \Gamma \tilde{\lambda} + 5 \tilde{\lambda}^2)\bigg{)}.
\end{aligned}
\end{equation}
\end{widetext}

As we see, $\chi(\tilde{\lambda})$ is exactly divisible by $P(\tilde{\lambda}) = \qty (\Gamma + \tilde{\lambda}) \qty ( \tilde{\lambda}^2 + 
 2 \Gamma \tilde{\lambda} -\frac{\tilde{\gamma} \Gamma}{2} + \frac{3 \Gamma^2}{4} )$. For the quotient of the division 
\begin{equation}\label{eq:quotient-division-polynomial}
\frac{\chi(\tilde{\lambda})}{P(\tilde{\lambda})} = a_0(\Gamma)\tilde\lambda^6 + a_1(\Gamma)\tilde\lambda^5 +\dots + a_5(\Gamma)\tilde\lambda + a_6(\Gamma),
\end{equation}
where $a_0(\Gamma)\equiv 1$, 
we apply the Newton diagram technique~\cite{Moro2003} to obtain the leading-order corrections to the eigenvalues. First of all, we consider a complex polynomial equation in $\tilde{\lambda}$:

\begin{equation}
    p\!\left (\tilde{\lambda}, \Gamma\right) = a_0 \left(\Gamma \right) \tilde{\lambda}^n  + \dots + a_{n-1}\left(\Gamma \right) \tilde{\lambda} + a_{n}\left(\Gamma \right) = 0,
\end{equation} with coefficients 

\begin{equation}\label{eq:Newton-diagram-coef-expansion}
    a_k \!\left(\Gamma\right) = \hat{\alpha}_k \Gamma^{\beta_k} + \dots, \quad k = 0, \dots, n, 
\end{equation}

where $\beta_k$ is the \emph{leading exponent} and $\hat{\alpha}_k$ the \emph{leading coefficient} of $\alpha_k(\Gamma)$. We set $a_0(\Gamma) \equiv 1$, i.e. $\hat{\alpha}_0 = 1$ and $\beta_0 = 0$.

Using Eq.~\eqref{eq:Newton-diagram-coef-expansion} we expand $a_k(\Gamma)$ in Eq.~\eqref{eq:quotient-division-polynomial} to leading order in $\Gamma$ and list the leading order terms in Table~\ref{tab:coefficients}.

\begin{table}[h!]
 \caption{Leading-order terms of the polynomial coefficients of $\chi(\tilde\lambda)/P(\tilde{\lambda})$.}
    \begin{tabular}{|l|l|l|l|l|l|l|}\hline
         $a_0(\Gamma)$ & $a_1(\Gamma)$ & $a_2(\Gamma)$ & $a_3(\Gamma)$ & $a_4(\Gamma)$ & $a_5(\Gamma)$ & $a_6(\Gamma)$\\ \hline
         $1\vphantom{\cfrac{\Gamma^2}{\Gamma^2}}$ & $6 \Gamma$ & $-\frac{5 \tilde{\gamma}}{2} \Gamma$ & $-\frac{\tilde{\gamma}^2}{2} \Gamma$ 
         & $-\frac{\tilde{\gamma}^2}{2} \Gamma^2$ & $-\frac{15 \tilde{\gamma}^4}{32} \Gamma$ & $-\frac{\tilde{\gamma}^4}{2} \Gamma^2$ \\\hline
    \end{tabular}
    \label{tab:coefficients}
\end{table}

\begin{figure}[t]
    \centering
    \includegraphics[width=.9\linewidth]{ 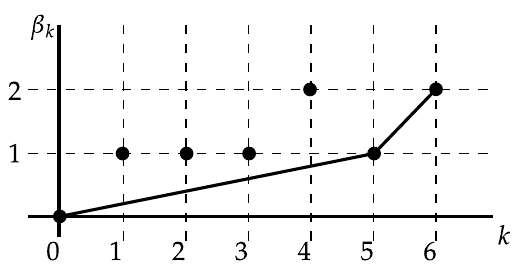}
    \caption{Newton polygon constructed from the leading-order coefficients of the characteristic polynomial in $\Gamma$ and $\tilde{\lambda}$.}
    \label{fig:Newton_polygon}
\end{figure}

Using the values $(k, \beta_k)$, we construct the Newton diagram for the quotient, which is displayed in Fig.~\ref{fig:Newton_polygon}. The Newton diagram is constructed to be the lower boundary of the convex hull of the set of points $\{ (k, \beta_k) \mid a_k \neq 0\}_{k = 0}^{n}$.
In our case, the diagram consists of two segments: one with slope $\frac{1}{5}$,
and one with slope $1$.
We can read the leading order exponents of the eigenvalue corrections and their multiplicities directly from the Newton diagram: the slopes of the segments give the exponents, while the horizontal spans of the segments --- the multiplicities. In our case, we obtain five eigenvalues that behave like $\propto \Gamma^{1/5}$ and one linear in $\Gamma$.

In order to obtain the leading coefficients $\mu$ of the eigenvalues of order $\Gamma^{\beta_k}$, we solve the characteristic equation
\begin{equation}
\sum_{k \in I_s} \mu^{n-k} \hat{\alpha}_k = 0.
\end{equation}
The set $I_s = \{ k : (k, \beta_k) \in S \}$ contains the indices of the points lying on a specific segment $S$ of the Newton diagram.

More precisely, the coefficients at the five eigenvalues of order $\Gamma^{\frac{1}{5}}$ are the nonzero roots of 
\begin{equation}
    \hat{\alpha}_0 \mu^6 + \hat{\alpha}_5 \mu = \mu^6 - \frac{15}{32} \tilde{\gamma}^4 \mu = 0,
\end{equation}
i.e., 
$
\mu = \left(\frac{15}{32}\right)^{1/5} \tilde{\gamma}^{4/5} e^{i \frac{2 \pi r}{5}}, \quad r = 1, \ldots, 5.
$
The coefficient for the only eigenvalue of linear order is the root of
\begin{equation}
\hat{\alpha}_5 \mu + \hat{\alpha}_6 = - \frac{15}{32} \tilde{\gamma}^4 \mu - \frac{1}{2} \tilde{\gamma}^4 = 0,
\end{equation}
\begin{equation}
    \mu = -\frac{16}{15}.
\end{equation}

These results are summarized as follows, restoring the original variables $\lambda$ and $\frac{\gamma_h - \gamma_e}{2}$ instead of $\tilde{\lambda}$ and $\tilde{\gamma}$:

\begin{itemize}
    \item For the block $J_5 \qty ( \lambda_{\mathcal{L}'_{\mathrm{eff}}})$:  $\lambda_{1 \dots 5} \approx  \lambda_{\mathcal{L}'_{\mathrm{eff}}} \!+ \!\qty (\frac{15}{512} \, \Gamma )^{1/5} \qty ( {\gamma}_{h} \!-\! {\gamma}_{e} )^{4/5} e^{i \frac{2 \pi r }{5} }$, for $r = 1 \dots5$  
    \item For the block $J_3 \qty (\lambda_{\mathcal{L}'_{\mathrm{eff}}})$: $\!\lambda_{6}\!=\!  \lambda_{\mathcal{L}'_{\mathrm{eff}}}-\!\Gamma\!$, \newline
    $\lambda_{7,8} \!=\!  \lambda_{\mathcal{L}'_{\mathrm{eff}}} \!- \!\Gamma \!\pm\! \frac{1}{2} \sqrt{\Gamma \qty (\Gamma + \gamma_h  - \gamma_e )} $ 
    \item For the block $J_1 \qty ( \lambda_{\mathcal{L}'_{\mathrm{eff}}})$: $\lambda_{9} \!\approx\!  \lambda_{\mathcal{L}'_{\mathrm{eff}}}\!-\!\frac{16}{15} \Gamma$
\end{itemize}

For the EP5 block \( J_5 \), a ring of five distinct simple eigenvalues appears, as described for the general case in Refs.~\cite{lidskii_perturbation_1966,moro_lidskii--vishik--lyusternik_1997}.

\section{Eigenvalues and the critical values of $\Omega_{ie}$ for the $\BL_\mathrm{eff}^{(ii)}$ in Sec.~\ref{sec: quantum jumps qutrit}}\label{sec:appE_eigenvals_qubit}

In the notation shown in Fig.~\ref{fig:qutrit_eigenvalues_spectrum}, the eigenvalues of the effective qutrit are ordered as follows

    \begin{align*}
        \lambda_1 &= -\gamma_{i}\\
        \lambda_2 &= -\gamma_{i}\\
        \lambda_3 &= - 2 \Gamma - \gamma_{i}\\
        \lambda_4 &= -\Gamma - \gamma_{i} - \sqrt{\Gamma^2 + \qty(\frac{\gamma_{h} - \gamma_{e}}{2})^2 - 8 \Omega_{ie}^2}\\
        \lambda_5 &= -\Gamma - \gamma_{i} + \sqrt{\Gamma^2 + \qty(\frac{\gamma_{h} - \gamma_{e}}{2})^2 - 8 \Omega_{ie}^2} \\
        \lambda_6 &= -\Gamma - \gamma_{i} - \frac{1}{2} \sqrt{4 \Gamma^2 + \qty(\frac{\gamma_{h} - \gamma_{e}}{2})^2 - 8 \Omega_{ie}^2}        
    \end{align*}
\newline
    \begin{align}
    \lambda_7 &= -\Gamma - \gamma_{i} - \frac{1}{2} \sqrt{4 \Gamma^2 + \qty(\frac{\gamma_{h} - \gamma_{e}}{2})^2 - 8 \Omega_{ie}^2}\nonumber\\
        \lambda_8 &= -\Gamma - \gamma_{i} + \frac{1}{2} \sqrt{4 \Gamma^2 + \qty(\frac{\gamma_{h} - \gamma_{e}}{2})^2 - 8 \Omega_{ie}^2}\\
        \lambda_9 &= -\Gamma - \gamma_{i} + \frac{1}{2} \sqrt{4 \Gamma^2 + \qty(\frac{\gamma_{h} - \gamma_{e}}{2})^2 - 8 \Omega_{ie}^2} \nonumber
    \end{align}

Then $\Omega_{ie}^{EP, \mathcal{L}_{eff}^{\prime}} = \frac{\gamma_{h} - \gamma_{e}}{4 \sqrt{2}}$, $\Omega_{ie}^{EP_{1}, \mathcal{L}_{eff}} = \frac{\sqrt{\Gamma^2 +  \qty(\frac{\gamma_{h} - \gamma_{e}}{2})^2}}{2 \sqrt{2}}$, $\Omega_{ie}^{EP_{2}, \mathcal{L}_{eff}} = \frac{\sqrt{4\Gamma^2 + \qty(\frac{\gamma_{h} - \gamma_{e}}{2})^2}}{2 \sqrt{2}}$.

\bibliography{main}

\end{document}